**Comparison of generative algorithms for conceptual groundwater modeling of coastal volcanic aquifer features with disparate, sparse and extremely imbalanced data**

Michael J. Friedel[1,2,3]

[1] Department of Physics, University of Colorado – Denver, USA

[2] Earthquest Consulting Ltd, Auckland, NZ

[3] Semeion Institute, Rome, IT

michael.friedel@ucdenver.edu

**ABSTRACT**

In coastal volcanic aquifers, the reliability of freshwater–seawater exchange simulations are governed by accuracy of the conceptual groundwater model (CGM). Traditional CGMs are constructed by qualitatively combining independent hydrogeophysical features, limiting their ability to capture the complexity of volcanic terrains. To integrate these disparate, sparse, and imbalanced features, we propose an AI-assisted workflow. First, the self-organizing map (SOM) is applied to estimate a deterministic set of transdisciplinary features called the *reference model*. Second, generative algorithms are applied to the reference model and empirical distributions constructed to obtain sets of stochastic point clouds called the *site model*. Data quality metrics identify the preferred generative algorithm whose set of stochastic features are mapped using SOM to the groundwater model grid and assigned as the *stochastic CGM*. The proposed algorithm is applied to extremely imbalanced multiclass features and multiple discrete numerical features observed at the Hālawa-Moanalua (H-M) aquifer, Oʻahu, Hawaiʻi. At this stie, the Copula Generative Adversarial Network is deemed as the preferred generative algorithm whose set of stochastic transdisciplinary features represent the H-M CGM. The simulated spatial geologic units correspond to published surface maps; and the simulated conductance, temperature, and barometric pressure profiles correlate with those measured at deep monitoring wells. Inspecting the 3D conductance models reveal groundwater flow and discharge driven by the aquifer's hydraulic gradient, freshwater pumping, seawater intrusion induced by onshore withdrawals, and preferred pathways for freshwater-seawater exchange, such as landward intrusion of seawater and seaward discharge of freshwater.

*Keywords:* Conditional tabular generative adversarial network - CTGAN, Copula generative adversarial network – CopulaGAN, Extremely-imbalanced data, Tabular Gaussian copula – TGC, Tabular variational autoencoder - TVAE, Transdisciplinary reference model, Stochastic transdisciplinary conceptual groundwater model.

**Highlights**

- AI-assisted CGM integrates sparse and highly imbalanced hydrogeophysical data.
- SOM and generative AI resolve complexity in coastal volcanic aquifers.
- CopulaGAN is statistically preferred for stochastic CGM development.
- 3D models reproduce surface geology and deep monitoring well feature profiles.
- Models reveal freshwater–seawater exchange and intrusion pathways.

## 1. Introduction

Volcanic islands are critical sources of freshwater for millions of people, yet their complex geology and topography pose significant challenges for groundwater exploration and management (Jenson et al., 2006; Gingerich and Engott, 2012; White et al., 2020). Freshwater in these environments is often stored in highly heterogeneous aquifers, including perched layers, basal lenses, and dike-impounded reservoirs, which are strongly influenced by lava flows, fractures, and volcanic structures (Stearns and Macdonald, 1946; Hunt, 1996; Thomas et al., 2019). Understanding the structure, storage, and dynamics of these aquifers is essential for sustainable water supply, especially in the context of climate change, sea-level rise, and increasing water demand across the Pacific region (Oki, 2004; Terry and Falkland, 2010; Rotzoll and Fletcher, 2013). Despite their importance, volcanic aquifers remain poorly characterized in many island settings due to the limited availability of subsurface data and the high variability of volcanic terrain (Custodio, 2010; Violette et al., 2012; Ingebritsen et al., 2015).

Traditionally, volcanic aquifer studies rely on borehole data, spring discharge measurements, and basic hydrological modeling to infer aquifer geometry, recharge rates, and freshwater–saltwater distribution (Stearns & Macdonald, 1946; Hunt, 1996; Custodio, 2010; Ahmed et al., 2018). While these approaches provide foundational insights, they are spatially limited and often unable to resolve the complex internal structure of volcanic formations (Thomas et al., 2019). Over the past couple decades, the application of geophysical methods transforms how to investigate and conceptualize volcanic aquifers (Violette et al., 2012; Baud et al., 2025). Common surface geophysical methods include time-domain electromagnetic (Alshehri and Abdelrahman, 2022; Carrasco-García et al., 2022), magnetotelluric (Hogg et al., 2018; Liu et al., 2022; Isaia et al., 2025), and magnetic resonance sounding (Legchenko et al., 2019).

Other researchers combine geophysical methods with the aim of reducing ambiguities in conceptual models of volcanic aquifers. Examples include combining electrical resistivity tomography and self-potential (Revil et al., 2008), airborne electromagnetic and magnetic data (Dumont et al., 2021), and electrical resistivity tomography and gravity (Khattach et al., 2025). Despite the growing number of geophysical studies in volcanic rock, few studies attempt to conceptualize subsurface systems from the mountains to the sea (Billy et al., 2022). Recently, Miller and Tontini (2025) developed a land-to-sea volcanic facies model by spatial clustering of density, susceptibility, and resistivity models derived through separate gravity, magnetic and airborne electromagnetic inversions.

In Hawai'i, sparse borehole sampling and disparate surface geophysical investigations infer the structural and hydrological complexity of volcanic aquifers. Geophysical methods, such as gravity, electrical resistivity, electromagnetic, magnetic, and seismic methods, identify dike-impounded groundwater systems, perched aquifers, and extensive basal freshwater lenses (Zohdy and Jackson, 1969; Gingerich and Engott, 2012; Izuka et al., 2015; Izuka et al., 2018; Thomas et al., 2019; Izuka and Rotzoll, 2023). On Hawai'i Island, marine controlled source electromagnetic and seismic studies reveal freshwater that extends offshore beneath the seafloor, possibly forming submarine groundwater reservoirs (Attias, et al., 2020). Other geophysical studies on the east side of Hawai'i Island use induced polarization measurements to determine surface temperature at the Kilauea shield volcano (Revil et al., 2021), and magnetotelluric measurements to image Kilauea's internal structure in terms of electrical resistivity (Hoversten et al., 2022).

On the island of O'ahu, surface geophysical studies are undertaken to conceptualize subsurface structures that may control flow through volcanic rocks in subregional aquifers at the Kaiwi Coastal and Hālawa-Moanalua areas. In the Kaiwi coastal area, seismic ambient noise surface wave tomography and self-potential results are independently used to infer groundwater flow in paleo-channels and along ridges of the basaltic bedrock (Grobbe et al., 2021). These qualitative inferences are nonunique and uncertain given the lack of direct borehole and/or water well information available to constrain their interpretation. In the Red Hill area, the subregional Hālawa-Moanalua aquifer spans the region from the upland dike complex (Walker, 1986) to low lands at the Pacific Ocean. In the central part of this aquifer, the seismic (Liberty and St. Clair, 2018) and gravity (Ito et al., 2019) survey results describe a plausible, albeit deterministic, and qualitative structural framework that infers the possible direction of groundwater flow. In this part of the aquifer, there also are borehole hydrogeologic observations (e.g., geology, water levels, and water quality) available to potentially constrain conceptualization of this volcanic aquifer. Given the disparate, sparse and imbalanced nature of this multidisciplinary data set, the traditional independent (visual) approach for feature integration is impractical and fraught with technical challenges, such as how to produce an unbiased model, estimate aquifer features and their uncertainty at nodes across a numerical groundwater model grid.

The lower third of the Hālawa–Moanalua aquifer reflects the coastal system and therefore may be hydraulically connected to secondary high-permeability pahoehoe layers with preferential flowpaths influenced by the Pacific Ocean, as on Hawai'i Island (Geng and Michael, 2020). Aquifer heterogeneity and preferential flow pathways are known to increase the risk of freshwater contamination (Geng and Michael, 2021), as spatial variability in hydraulic conductivity govern seawater intrusion by altering the freshwater–seawater interface through mixing and spreading (Adams and Gelhar, 1992; Dentz et al., 2011). Numerical groundwater models demonstrate that connected heterogeneities can generate complex offshore salinity patterns and enhance seawater intrusion under pumping (Siena and Riva, 2018; Yu and Michael, 2019). In fractured volcanic aquifers, highly permeable conduits, such as those in pahoehoe lava flows, may serve as preferential pathways that exacerbate seawater intrusion threatening freshwater resources (Bonacci and Roje-Bonacci, 1997; Xu et al., 2016). Three-dimensional numerical simulations reveal that conduit geometry and connectivity can strongly influence the extent and variability of seawater intrusion, underscoring the need to represent 3D heterogeneity and preferential flowpaths in coastal groundwater models. Fully resolving these processes in 3D models, however, remains challenging. Primary challenges include poorly resolved conceptual groundwater models, and numerical models that are computationally expensive and prone to numerical instability (Yu and Michael, 2019).

Current trends in Earth, energy and environmental studies use artificial intelligence (AI) to inform problem-solving and decision-making. AI applications in groundwater hydrology include data-driven machine learning, physics-informed machine learning, and multimodal machine learning. Data driven machine learning largely focuses on supervised machine learning methods to map and forecast groundwater levels, as reviewed by De Salvo et al. 2022. Physics-informed machine learning brings together supervised data driven machine learning with differential equations to forecast space-time groundwater flow and/or transport (Yeung et al., 2022; Friedel and Buscema, 2023; Adombi et al., 2024). Multimodal machine learning affords the possibility to incorporate any number of categorical and numerically derived values along with point field measurements to inform the CGM. These features can be derived using numerical model calibrations/inversions, supervised and unsupervised data-driven, physics-informed, and deep learning approaches. Recent work by University of Hawai'i researchers demonstrates the applicability of multimodal machine learning for developing a CGM reflecting groundwater-geothermal interaction and their features across Lanai and Hawai'i Islands (Friedel et al., 2023).

The aim of this study is to develop and evaluate an AI-assisted workflow for developing CGMs of island volcanic aquifers. This study addresses the research question: Can generative statistical and/or generative AI algorithms be used to characterize Pacific volcanic groundwater system features using disparate, sparse and extremely imbalanced data? Answering this question is important given that the quality of a numerical groundwater model is limited by adequacy of the CGM. To answer this question, we hypothesize that mutual information (Shannon, 1948) among hydrogeophysical (geologic, geophysical, engineering, and water quality) observations can be computationally assimilated and used to produce a valid stochastic transdisciplinary set of aquifer features representing the CGM. In testing this hypothesis, an AI-assisted workflow is proposed for application to volcanic aquifers. The AI-assisted workflow is tested to achieve three objectives at the sub-regional Hālawa-Moanalua (H–M) aquifer on O'ahu, Hawai'i: (1) compare performance of the unsupervised machine learning to provide deterministic estimates of transdisciplinary features; (2) compare performance of the different generative algorithms in predicting alternate transdisciplinary feature sets; and (3) use the preferred generative algorithm to quantify, map, and interpret the 3D stochastic distribution of geological, geophysical, engineering, and water-quality features. This study extends the work of Friedel et al. (2023), who developed a deterministic transdisciplinary data-driven model of freshwater-geothermal system features at selected locations in actual and synthetic boreholes on Lanai and Hawai'i Islands. This study also expands the work of researchers who applied the Gaussian Copula to generate groundwater quality parameters for training a deep neural network for estimation and uncertainty analysis of a coastal aquifer experiencing seawater intrusion (Taşan et al., 2023; Zhang et al., 2024).

## 2. Methods

The reliability of freshwater–seawater exchange simulations are governed by accuracy of the conceptual groundwater model (CGM). Traditional CGMs are constructed by qualitatively combining independent hydrogeophysical features, limiting their ability to capture the complexity of volcanic aquifers. Integrating features using this ad-hoc approach is difficult because categorical and numerical observations oftentimes are disparate, sparse, skewed, and extremely imbalanced. To overcome these limitations, we propose the following quantitative AI-assisted workflow: the Data cube, the Reference Model, the Site Model, and the Conceptual Groundwater Model (Fig. 1). Each of these steps are briefly described next.

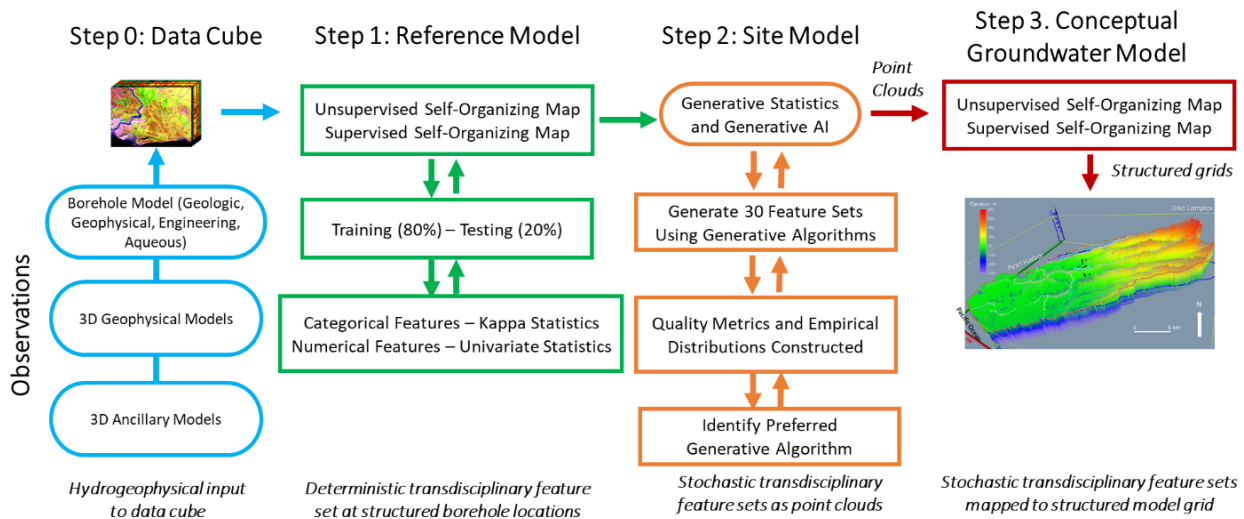

**Fig. 1.** Proposed AI-assisted workflow to improve reliability of Conceptual Groundwater Models across Pacific Island volcanic aquifers.

*2.1. Data cube*

Development of the AI-assisted workflow begins with selecting a Pacific Island volcanic aquifer where diverse hydrogeophysical observations exist and/or will be gathered. To enable system-wide analysis, all of the available and future observations are compiled into an unstructured data cube. The data cube provides a unified framework for aggregating categorical and numerical features across spatial location, depth, and data type. This structured framework supports the joint analysis of direct and indirect observations.

Direct observations can include physical, chemical, and physicochemical measurements collected from wells, boreholes, and ocean stations. Indirect observations are derived primarily from geophysical methods—such as electrical resistivity, electromagnetic, gravity, and seismic surveys—as well as drilling-based interpretations, equations, and published datasets. These hydrogeophysical observations can be represented as categorical (discrete) and numerical (continuous) features distributed across heterogeneous and unstructured grid nodes. Integration across data sources is achieved through self-organization of mutual information (Shannon, 1948; Kraskov et al., 2004) in hydrogeophysical features.

Categorical features may include geologic units, aquifer confinement, and open-ocean. These variables will be identified using one-hot encoding (present = 1, absent = 0, unknown = blank), converting categorical labels into a numerical format suitable for machine learning (Google Developers, 2025). Continuous features consist of both direct measurements and indirectly derived parameter estimates, including geophysical properties and engineering metrics. All categorical and continuous variables incorporated into the data cube are treated as soft priors, with uncertainty evaluated during generative AI–based inference through shared information among reference model features (Kraskov et al., 2004). The combined dataset may exhibit class imbalance in geologic units and spatial skewness in numerical variables due to uneven sampling and heterogeneous hydrogeologic conditions. To mitigate these effects, preprocessing steps will include sample augmentation and logarithmic transformation of selected numerical variables, providing the foundation for the information-driven generative modeling framework described in the following section.

*2.2. Reference model*

*2.2.1 Self organizing map*

The reference model is constructed using a two-stage self-organizing map (SOM) framework that combines unsupervised pattern discovery with supervised estimation of aquifer features. SOMs are neural network–based, unsupervised learning algorithms that project high-dimensional data onto a low-dimensional (typically two-dimensional) lattice while preserving the topological relationships of the input space (Kohonen, 2001, 2013). This property makes SOMs well suited for exploratory analysis and dimensionality reduction of complex hydrogeological datasets, where nonlinear dependencies and multivariate correlations are common.

In the first stage, an unsupervised SOM is trained to learn the intrinsic structure of the multivariate aquifer data extracted from the data cube (after Friedel et al., 2020). Training follows a competitive learning procedure in which input vectors are iteratively presented to the network and mapped to the best-matching unit (BMU), defined as the node whose weight vector minimizes the Euclidean distance to the input vector,

$$D = \sqrt{\sum_{i=0}^{N}(V_i - W_i)^2}$$

where **V** is the input vector and **W** is the weight vector. A neighborhood of nodes surrounding the BMU is identified using a Gaussian kernel with a time-dependent radius,

$$\sigma = \sigma_0 exp\left(-\frac{1}{\lambda}\right), t = 1,2,3, \ldots,$$

where $\sigma_0$ is radius at $t_0$, $\lambda$ is a decay constant, and t denotes the iteration. The weights of the BMU and its neighbors are updated according to

$$\mathbf{W}(t+1) = \mathbf{W}(t) + L(t)(V(t) - \mathbf{W}(t)),$$

where the learning rate $L(t) = L_0 exp\left(-\frac{1}{\lambda}\right), t = 1,2,3, \ldots$, decays exponentially over time, and the neighborhood influence function is given by

$$\Theta(t) = exp\left(-\frac{dist^2}{2\sigma^2(t)}\right), t = 1,2,3, \ldots,$$

Here, *dist* denotes the distance between a node and the BMU on the SOM grid. Competitive training proceeds iteratively until changes in the weight vectors converge, yielding a topologically ordered representation of the aquifer feature space.

In the second stage, a supervised SOM is applied to associate the trained SOM nodes with target variables representing aquifer properties of interest (Riese et al., 2020). Continuous variables are predicted using node-based regression, while categorical variables are assigned probabilistic class memberships based on the distribution of labeled samples mapped to each node. The combined regression and classification outputs define a deterministic, data-driven reference model that captures the dominant multivariate structure of the aquifer system and serves as a baseline for subsequent uncertainty quantification and synthetic data generation analyses.

*2.2.2. Training and testing*

Model training and evaluation are conducted using k-fold cross-validation (Hastie et al., 2009), in which the dataset is randomly shuffled and divided into training (typically 80%) and testing (20%) subsets that are repeated *N* times for assessing model generalizability to independent field data. Model performance is evaluated using metrics appropriate for continuous and categorical features, e.g., the coefficient of determination ($R^2$) and mean squared error (MSE) for continuous variables, and overall accuracy and Cohen's kappa statistic (Cohen, 1960) for categorical variables. In k-fold cross-validation, the dataset is partitioned into *k* equal subsets, with the model trained on *k – 1* folds and validated on the remaining fold, repeated until each fold is used once for validation, and performance metrics averaged to reduce bias and maximize data use. Cohen's kappa measures agreement beyond chance and is interpreted as ≤ 0 (no agreement), 0.01–0.20 (none to slight), 0.21–0.40 (fair), 0.41–0.60 (moderate), 0.61–0.80 (substantial), and 0.81–1.00 (almost perfect), providing a robust assessment of classification reliability in ML-based digital geologic mapping.

*2.2.3. Grouping features*

Statistically meaningful feature grouping of self-organized information is performed through k-means clustering (Vesanto and Alhoniemi, 2000) applied to the SOM codebook vectors across the trained network. Optimal partitioning for each candidate number of clusters is determined using the Euclidean distance criterion, with cluster validity evaluated using the Davies–Bouldin index (Davies and Bouldin, 1979). To reduce sensitivity to local minima, the k-means algorithm is repeated multiple times for each clustering configuration. Although the Davies–Bouldin index is used here for convenience, other cluster validity measures could be applied. The final reference model is constructed by retraining the SOM with optimized parameters using the complete dataset (combined training and testing observations) from the data cube.

*2.3 Site model*

*2.3.1. Synthetic data generation methods*

The *site model* is constructed through application of synthetic data generation methods to the reference model. Synthetic data generation is employed to produce stochastic ensembles of transdisciplinary aquifer features for use in conceptual groundwater modeling and uncertainty quantification. Four complementary methods are evaluated representing a range of statistical assumptions and modeling capacities: a Tabular Gaussian Copula (TGC), a Tabular Variational Autoencoder (TVAE), a Conditional Tabular Generative Adversarial Network (CTGAN), and a Copula Generative Adversarial Network (CopulaGAN). These approaches span classical statistical, variational, and adversarial learning frameworks, each offering distinct trade-offs in interpretability, computational efficiency, and ability to capture complex dependencies in heterogeneous hydrogeological datasets.

The TGC models the joint distribution of a multivariate dataset by separating marginal behavior from dependence structure (Nelsen, 2006). Each variable $X_j$ is transformed to uniform space using its marginal cumulative distribution function (CDF), $U_j = F_j(X_j)$, and mapped to a Gaussian latent space via $Z_j = \Phi^{-1}(U_j)$, where $\Phi^{-1}$ is the inverse standard normal CDF. Dependence is captured through a correlation matrix $\Sigma$ estimated from $\mathbf{Z} \sim \mathcal{N}(0, \Sigma)$, and synthetic samples are generated by sampling $\mathbf{Z^*} \sim \mathcal{N}(0, \Sigma)$ and applying inverse marginal transformations $X_j^* = F_j^{-1}(\Phi(Z_j^*))$. This approach is computationally efficient and transparent, making it well suited for baseline uncertainty characterization, though it is limited to linear dependence structures.

The TVAE extends variational autoencoders to mixed continuous–categorical tabular data by learning a probabilistic latent representation through an encoder–decoder architecture (Kingma and Welling, 2014; Xu et al., 2019). The encoder defines an approximate posterior $q_\phi(z \mid x) = \mathcal{N}(\mu_\phi(x), \text{diag}(\sigma_\phi^2(x)))$, while the decoder models $p_\vartheta(x \mid z)$ using variable-specific likelihoods. Model training maximizes the evidence lower bound,

$$\mathcal{L}_{ELBO} = \mathbb{E}_{q\phi(z|x)} = [log p_\Theta(\boldsymbol{x} \mid \boldsymbol{z})] - KL\left(q_\phi(\boldsymbol{z} \mid \boldsymbol{x}) \parallel p(\boldsymbol{z})\right),$$

with $p(z) = \mathcal{N}(0, I)$. TVAE enables modeling of nonlinear multivariate relationships but may struggle with rare events and sharp distributional features.

The CTGAN addresses limitations of conventional GANs for tabular data by introducing conditional sampling and specialized preprocessing for mixed data types and imbalanced categories (Goodfellow et al., 2014; Xu et al., 2019). The generator produces synthetic samples conditioned on a categorical vector $c$, $x^* = G_\vartheta(z, c)$, where $z \sim \mathcal{N}(0, I)$. Training commonly adopts a Wasserstein GAN with gradient penalty (WGAN-GP), with discriminator loss

$$\mathcal{L}_D = \mathbb{E}_{x \sim p_{data}} =[D_\phi(x,c)] - \mathbb{E}_{\tilde{x} \sim pG}\left[D_\phi(\tilde{x},c)\right] + \lambda \mathbb{E}_{\hat{x}}\left[\|\nabla_{\hat{x}} D_\phi(\hat{x},c)\|_2 - 1\right]^2,$$

and generator loss $\mathcal{L}_G = -\mathbb{E}_{\tilde{x} \sim pG}\left[D_\phi(\tilde{x},c)\right]$. CTGAN is well suited for large, heterogeneous datasets but is computationally intensive and sensitive to hyperparameter choices.

The CopulaGAN combines copula-based marginal transformation with adversarial learning in copula space to preserve marginal distributions while learning nonlinear dependence structures (Torbenson et al., 2020). Observations are first mapped to copula space via $u_j=F_j(x_j)$, and a GAN is trained to generate synthetic copula samples $\tilde{u} = G_\theta(z)$. These samples are mapped back to the original data space using inverse marginals, $x_j^* = F_j^{-1}(\tilde{u}_j)$. This hybrid approach improves marginal fidelity and training stability relative to raw GANs, though it remains less mature than alternative tabular methods.

The comparison of these four generative methods is motivated by the need to support data assimilation and uncertainty quantification in groundwater modeling, where accurate representation of subsurface variability and inter-parameter dependence strongly influences predictive uncertainty. Copula-based models provide statistically grounded baselines for uncertainty propagation, while variational and adversarial models offer greater flexibility for capturing nonlinear hydrogeological relationships. Evaluating multiple generative paradigms allows assessment of how structural modeling assumptions affect stochastic aquifer feature ensembles, reducing reliance on a single representation of subsurface heterogeneity and strengthening the robustness of uncertainty estimates in subsequent groundwater simulations.

*2.3.2. Validation of tabular generative algorithms*

The performance of synthetic data generation models is evaluated using the standardized, model-agnostic metric framework provided by the Synthetic Data Vault (SDV), which enables consistent comparison of generative algorithms based on statistical fidelity and preservation of multivariate dependencies (Patki et al., 2016). These quality metrics quantify similarity between real and synthetic data by comparing univariate distributions—using nonparametric measures such as the Kolmogorov–Smirnov (KS) statistic for continuous variables and total variation distance for categorical variables—and by assessing the preservation of pairwise relationships between variables. These metrics are well suited to groundwater modeling studies affected by seawater intrusion, where maintaining realistic concentration distributions and dependencies among salinity, major ions, and hydrochemical indicators is critical. Unlike GAN-specific or low-dimensional goodness-of-fit metrics, the SDV metrics scale efficiently to high-dimensional datasets and support transparent, reproducible evaluation across statistical, machine-learning, and deep-learning models, consistent with established hydrological practice (Wilcox, 2017; Arndt et al., 2020).

The SDV framework organizes evaluation metrics into different types: Diagnostic, Quality, and Privacy. Diagnostic metrics verify data validity and structural integrity by checking data types, value ranges, categorical levels, table structure, and completeness, helping to identify invalid or failed synthetic outputs. Quality metrics form the primary basis for model comparison and measure both column-level distribution similarity and the preservation of relationships between variable pairs, including numerical, categorical, and mixed-variable interactions. These metrics are normalized between 0 and 1 and aggregated into an overall Quality Score. The SDV framework does not, by default, include downstream predictive performance metrics, domain-specific physical constraints, or coverage diagnostics; such measures must be defined separately when required. In this study, generative models are compared primarily using SDV Quality Scores, while privacy metrics are reported independently and are not used for ranking herein.

To evaluate the fidelity of synthetic tabular datasets generated using TGC, TVAE, CTGAN, and CopulaGAN models, the KS Complement metric is employed to quantify agreement between the marginal distributions of real and synthetic variables (Dankar, et al., 2025). The metric is derived from the Kolmogorov–Smirnov (*KS*) statistic, which measures the maximum absolute difference between the empirical cumulative distribution functions (CDFs) of observed data $F_{obs}(x)$ and synthetic data $F_{syn}(x)$, defined as $KS = \sup_x | F_{obs}(x) - F_{syn}(x) |$. The *KS* Complement is computed as $1 - KS$, yielding a normalized similarity score ranging from 0 to 1, with higher values indicating closer correspondence between observed and synthetic distributions. This transformation facilitates direct comparison across models and variables.

The KS Complement is calculated for each continuous hydrological variable independently and subsequently aggregated across variables to provide an overall measure of distributional fidelity. As a non-parametric statistic, the metric does not impose assumptions on the underlying data distributions and is therefore well suited for the heterogeneous and often non-Gaussian characteristics of hydrological variables. However, because the KS statistic evaluates marginal distributions only and does not capture inter-variable dependence structures, it is interpreted alongside complementary metrics assessing correlation preservation and multivariate dependence to provide a comprehensive evaluation of synthetic data quality.

## 3. Results and discussion

*3.1 Data cube*

The first step in building an AI-assisted conceptual groundwater model (CGM) involves the selection of a study site from where geologic, geophysical, engineering, and water quality (referred to as hydrogeophysical) observations can be acquired and aggregated into the data cube. For this purpose, the Hālawa–Moanalua (H-M) sub-regional aquifer on the island of Oʻahu, Hawaiʻi, is selected (Fig. 2). The H-M aquifer spans six surface catchments: Aiea (5.30 km²), Hālawa (25.4 km²), Moanalua (27.2 km²), Salt Lake (19.0 km²), unnamed catchment #33030 (11.4 km²), and unnamed catchment #34030 (6.14 km²). This aquifer is bounded to the north by the upland dike complex (Walker, 1986) and to the south by the Pacific Ocean. The lower third of the study area is bounded by Pearl Harbor to the west, the Pacific Ocean to the south, and an unnamed harbor to the east (referred hereafter as the East Harbor). Under these conditions, precipitation entering at the dike complex is hypothesized to flow under hydraulic gradient toward the Pacific Ocean.

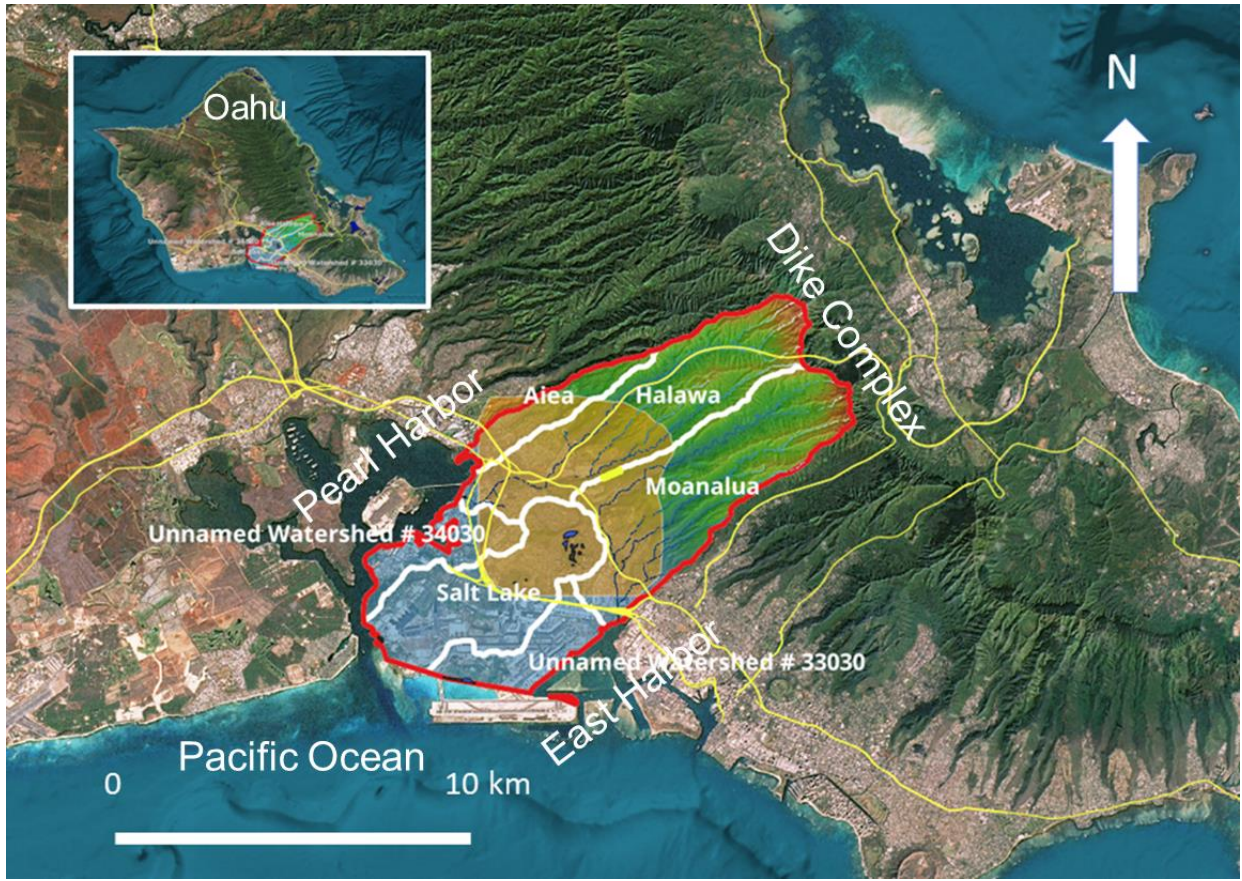

**Fig. 2.** Map of the study area, Oʻahu Hawaiʻi (inset), and primary catchments (outlined in white) overlying the Hālawa-Moanalua aquifer (outlined in red). Primary catchment areas are: Aiea (5.30 km$^2$), Hālawa (25.4 km$^2$), Moanalua (27.2 km$^2$), Salt Lake (19.0 km$^2$), unnamed #33030 (11.4 km$^2$), and unnamed #34030 (6.14 km$^2$).

Key datasets include borehole logs from Naval Facilities (NavFac, 2018–2021), water wells (Lautze, 2014a, b), deep monitoring wells and water supply wells (Board of Water Supply, 2023; U.S. Geological Survey, 2023), water quality wells (Lautze et al., 2014a, b; Lautze, 2017), harbor seawater profile nodes surrounding the aquifer (Lenntech, 2023; PacIOOS, 2023; PetroWiki, 2023), and synthetic wells generated as part of this study. A reference map summarizing the locations and observation types used to populate the data cube is shown in Fig. 3. Feature assignments integrate multiple complementary data sources, including direct borehole observations (NavFac, 2018–2021), seismic refraction and reflection data delineating subsurface contacts (Liberty and St. Clair, 2018), drilling depths to the caprock–pāhoehoe interface (Hunt, 1997), published surface geologic maps with unit elevations (Sherrod et al., 2021), harbor profile nodes (PacIOOS, 2023), deep well records (Board of Water Supply, 2023), and the synthetic wells developed herein. These key datasets comprise categorical and continuous features described next.

*3.1.1. Categorical features*

Categorical features, including geologic units ('aʻā, alluvium, caprock, loose and welded clinkers, pāhoehoe, saprolite, and tuff), aquifer confinement (confined, unconfined), and open-ocean (yes, no), are represented using one-hot encoding (Google Developers, 2025). Example categorical encodings presented below include: geologic units (positions 1-8), aquifer confinement (position 9), and open ocean (position 10).

- 'aʻā, confined aquifer, no ocean → [1,0,0,0,0,0,0,0,1,0]
- alluvium, confined aquifer, no ocean → [0,1,0,0,0,0,0,0,1,0]
- caprock, confined aquifer, no ocean → [0,0,1,0,0,0,0,0,1,0]
- clinker loose, unconfined aquifer, no ocean → [0,0,0,1,0,0,0,0,0,0]
- clinker welded, confined aquifer, no ocean → [0,0,0,0,1,0,0,0,1,0]
- saprolite, confined aquifer, no ocean → [0,0,0,0,0,1,0,0,1,0]
- pāhoehoe, confined aquifer, no ocean → [0,0,0,0,0,0,1,0,1,0]
- tuff, unknown confinement, no ocean → [0,0,0,0,0,0,0,1, ,0]
- unknown, unknown confinement, no ocean → [ , , , , , , , , ,0]

Other geologic units also are identified indirectly from velocity measurements derived from seismic refraction surveys (Liberty and St. Clair, 2018), which are translated into lithologic units using published velocity ranges summarized in Table 1. Where multiple velocities correspond to a single unit, median values are used. To improve spatial coverage and support data-driven inference, synthetic wells lacking assigned features (empty cells) are incorporated as placeholders within the Data Cube. These synthetic wells are distributed throughout the aquifer (white circles; Fig. 3) to depths of 600 m, with higher densities near key structural features north of the tuff cone and along seismic survey lines.

**Table 1.**
P-wave velocity values used for assigning geologic units for incorporation into the Reference Model.

| Geology | | Velocity, m/s | | | | |
|---|---|---|---|---|---|---|
| Type | Unit | Minimum | Median | Maximum | Comment | References |
| Alluvium | alluvium | 0 | 1000 | | Unconsolidated sediment deposited | Von Voigtlander et al., 2017 |
| | caprock | 3000 | 3100 | 3200 | Marine and terrestrial sediments | Von Voigtlander et al., 2017; Souza and Voss, 1987 |
| Basalt | pahoehoe | >3400 | 5000 | <=5300 | Lava flow | Bücker, 1999; Von Voigtlander et al., 2017 |
| | aa | >5500 | 5750 | <6000 | lava flow | Brandes et al., 2011; Von Voigtlander et al., 2017 |
| | clinker (loose) | >1900 | 2175 | 2450 | lava flow | Brandes et al., 2011; Von Voigtlander et al., 2017 |
| | clinker (welded) | 2450 | 2850 | <=3000 | lava flow | Visher and Mink, 1964; Von Voigtlander et al., 2017 |
| | saprolite | >=1000 | 1450 | <=1900 | Weathered in-place | Miller, 1988; Von Voigtlander et al., 2017 |
| Volcanic | tuff | >3000 | 3000 | <=3400 | volcanic ash and debris | Keller, 1960; Finstick, 1999; Von Voigtlander et al., 2017 |

Vertical assignment of geologic units follows a set of systematic rules. Saprolite is assigned to nodes extending 3 m above the saprolite–pāhoehoe contact, while pāhoehoe is assigned to nodes extending 3 m below this contact. Caprock is assigned from the caprock–pāhoehoe contact upward to the land surface, and pāhoehoe is assigned downward from this contact to the base of the model. These assignments assume undifferentiated caprock, saprolite, and pāhoehoe units. Surface occurrences of alluvium and tuff are assigned based on published geologic maps. A summary of categorical feature counts is provided in Table 2.

Categorical observations—including feature type, name, and Data Cube reference—are listed in Table 3. Examination of total counts and fractional representation reveals an extremely imbalanced dataset. For example, saprolite and pāhoehoe together account for more than half of all observations (saprolite: *n* = 25,310, 33%; pāhoehoe: *n* = 13,295, 18%), whereas several units are sparsely represented, including welded clinker ($n$ = 2,507, 3%), 'a'ā ($n$ = 2,580, 3%), tuff ($n$ = 3,082, 4%), and ocean ($n$ = 28, <0.1%). Caprock ($n$ = 3,802, 5%) and alluvium ($n$ = 12,083, 16%) occupy intermediate proportions, while hydrogeologic conditions are also unevenly represented (confined: $n$ = 2,737, 4%; unconfined: $n$ = 6,613, 9%).

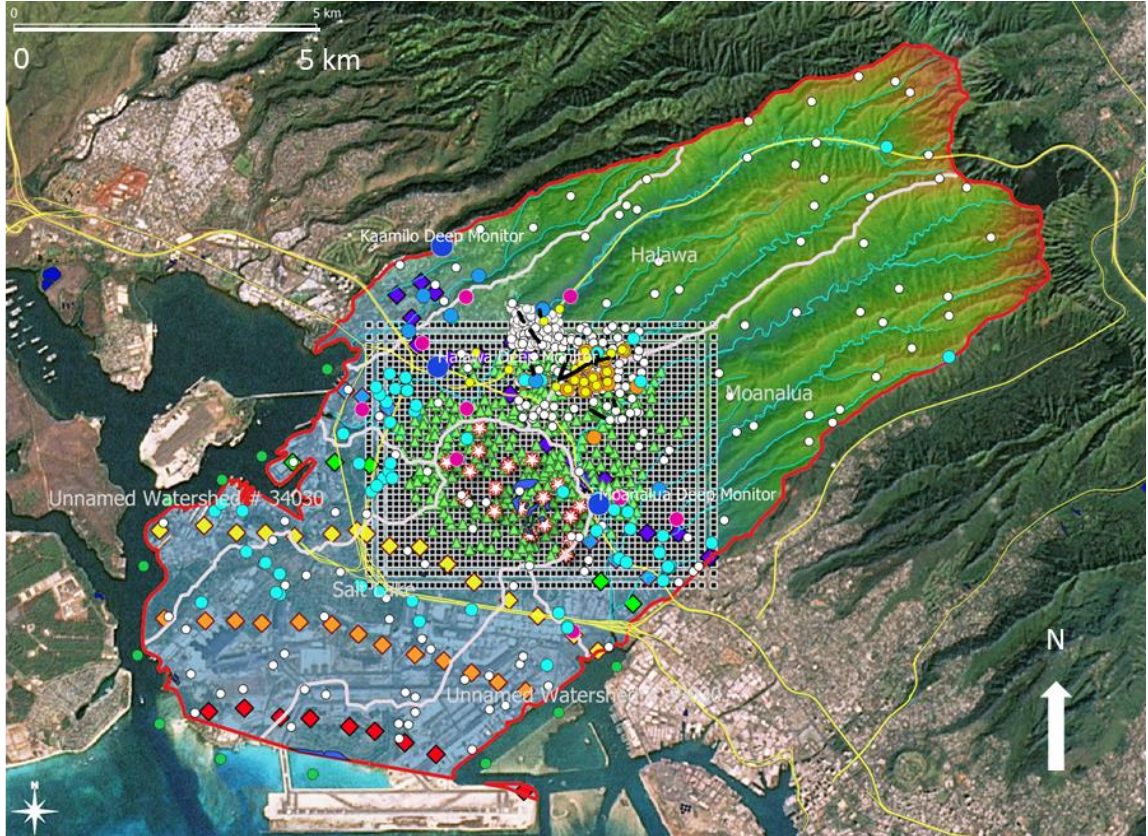

**Fig. 3.** Spatial distribution of hydrogeophysical observations used to construct the reference model for the Hālawa–Moanalua aquifer. Symbols indicate observation type and depth constraint. Triangles show interpreted depths to the caprock–pāhoehoe contact (red = 304 m, orange = 243 m, yellow = 183 m, green = 122 m, cyan = 61 m, blue = 0 m), and circles denote wells, including water wells (light blue), deep monitoring wells (royal blue), water supply wells (magenta), synthetic wells (white), and harbor nodes (green). Gravity stations are shown as green triangles. The right panel illustrates observation density within the central aquifer, including inverted geophysical properties (density from gravity and velocity from seismic refraction), interpreted seismic reflections marking lithologic contrasts, and hydrogeologic measurements collected at disposal, plume, and monitoring wells. Measured features include lithology, rock quality designation, fracture rate, oxidation–reduction potential, specific conductance, temperature, and water level.

*3.1.2. Continuous features*

Seismic velocity values within the Data Cube are derived using multiple complementary approaches: (i) conversion from rock quality designation (RQD) using published empirical relationships (Fathani et al., 2018); (ii) assignment of median velocities by geologic unit based on published ranges (Bücker et al., 1999; Miller et al., 2016); and (iii) inversion of seismic refraction data (Liberty and St. Clair, 2023). Depth-dependent seawater properties to approximately 25 m are obtained from harbor seismic profiles (PacIOOS, 2023). Water-level and water-quality data are sourced from monitoring wells (USGS, 2023) and the Hawaiʻi Play Fairway Analysis datasets (Lautze et al., 2014a, b; Lautze, 2017). Temperature, barometric pressure, and specific conductance data are obtained from deep monitoring wells maintained by the Board of Water Supply (2023). A summary of numerical feature counts incorporated into the reference model is provided in Table 3.

Indirect subsurface constraints are incorporated through geophysical models, including a three-dimensional density model derived from gravity data (Ito et al., 2019) and two-dimensional seismic refraction velocity and reflection sections along survey profiles (Liberty and St. Clair, 2018). Seismic reflections are interpreted as impedance contrasts at the saprolite–pāhoehoe contact. Above this interface, saprolite is assumed to be vertically homogeneous extending upward to the base of alluvium (where present), while undifferentiated pāhoehoe extends downward from the contact to the base of the model. The saprolite is further assumed to be homogeneous and impermeable, thereby hydraulically confining the underlying pāhoehoe within the study area.

Additional structural constraints include digitized interpretations of the caprock–pāhoehoe contact derived from drilling records (Hunt, 1997). At these locations, caprock extends from the contact to the aquifer surface, with the underlying undifferentiated pāhoehoe assumed to be confined and laterally continuous. Drill records are grouped into discrete depth intervals (0–304 m; Fig. 3), revealing an impermeable wedge that thickens seaward. A limited number of soft prior constraints are also incorporated from mapped surface occurrences of alluvium and tuff (Sherrod et al., 2021). Finally, additional velocity estimates are assigned to borehole geologic units by applying empirical RQD–velocity relationships to observed borehole RQD values (Fathani et al., 2018).

**Table 2.**
Summary of categorical feature counts included in the Data cube. The table includes observation (direct or indirect), type (borehole, seismic refraction, drill depth, map, harbor, deep wells, and synthetic wells), name of observation, categorical feature counts (geologic unit and aquifer confinement), and reference.

| | | | Categorical Feature Count | | | | | | | | | | | |
|---|---|---|---|---|---|---|---|---|---|---|---|---|---|---|
| | | | Geologic Unit | | | | | | | | Aquifer Confinement | | | |
| Observation | Type | Name | caprock | alluvium | saprolite | clinker (loose) | clinker (welded) | aa | pahoehoe | tuff | Confined | Unconfined | ocean | Total | Reference |
| Direct | Borehole | RHP01 | 0 | 5 | 5 | 5 | 41 | 79 | 25 | 24 | 49 | 135 | 0 | 368 | NavFac, 2018–2021 |
| Direct | Borehole | RHP02 | 0 | 5 | 0 | 28 | 14 | 99 | 22 | 0 | 49 | 119 | 0 | 336 | NavFac, 2018–2021 |
| Direct | Borehole | RHP03 | 0 | 5 | 14 | 15 | 24 | 85 | 19 | 0 | 49 | 114 | 0 | 325 | NavFac, 2018–2021 |
| Direct | Borehole | RHP04A | 0 | 24 | 3 | 28 | 33 | 69 | 8 | 0 | 49 | 140 | 0 | 354 | NavFac, 2018–2021 |
| Direct | Borehole | RHMW01R | 0 | 48 | 6 | 56 | 66 | 138 | 109 | 0 | 258 | 688 | 0 | 1369 | NavFac, 2018–2021 |
| Direct | Borehole | RHMW02 | 0 | 0 | 0 | 15 | 4 | 20 | 64 | 0 | 21 | 478 | 0 | 602 | NavFac, 2018–2021 |
| Direct | Borehole | RHMW03 | 0 | 0 | 0 | 20 | 6 | 28 | 34 | 0 | 30 | 520 | 0 | 638 | NavFac, 2018–2021 |
| Direct | Borehole | RHMW04 | 0 | 1 | 0 | 0 | 0 | 0 | 35 | 0 | 27 | 294 | 0 | 357 | NavFac, 2018–2021 |
| Direct | Borehole | RHMW05 | 0 | 5 | 0 | 22 | 0 | 29 | 0 | 0 | 20 | 363 | 0 | 439 | NavFac, 2018–2021 |
| Direct | Borehole | RHMW6 | 0 | 0 | 36 | 8 | 0 | 129 | 0 | 0 | 42 | 239 | 0 | 454 | NavFac, 2018–2021 |
| Direct | Borehole | RHMW7 | 0 | 0 | 5 | 34 | 0 | 0 | 0 | 0 | 48 | 193 | 0 | 280 | NavFac, 2018–2021 |
| Direct | Borehole | RHMW8 | 0 | 15 | 0 | 78 | 0 | 80 | 89 | 28 | 29 | 291 | 0 | 610 | NavFac, 2018–2021 |
| Direct | Borehole | RHMW09 | 0 | 2 | 10 | 40 | 3 | 160 | 168 | 24 | 38 | 376 | 0 | 821 | NavFac, 2018–2021 |
| Direct | Borehole | RHMW10 | 0 | 11 | 9 | 59 | 6 | 178 | 230 | 7 | 34 | 476 | 0 | 1010 | NavFac, 2018–2021 |
| Direct | Borehole | RHMW11 | 0 | 68 | 193 | 37 | 6 | 57 | 132 | 0 | 318 | 192 | 0 | 1003 | NavFac, 2018–2021 |
| Direct | Borehole | RHMW12A | 0 | 9 | 7 | 15 | 31 | 94 | 284 | 6 | 290 | 220 | 0 | 956 | NavFac, 2018–2021 |
| Direct | Borehole | RHMW13 | 0 | 22 | 22 | 92 | 2 | 128 | 263 | 2 | 301 | 230 | 0 | 1062 | NavFac, 2018–2021 |
| Direct | Borehole | RHMW14 | 0 | 41 | 60 | 32 | 10 | 51 | 302 | 0 | 350 | 160 | 0 | 1006 | NavFac, 2018–2021 |
| Direct | Borehole | RHMW15 | 0 | 2 | 3 | 81 | 62 | 177 | 262 | 3 | 300 | 290 | 0 | 1180 | NavFac, 2018–2021 |
| Direct | Borehole | RHMW16 | 0 | 40 | 37 | 42 | 24 | 93 | 263 | 0 | 312 | 198 | 0 | 1009 | NavFac, 2018–2021 |
| Direct | Borehole | RHMW17 | 0 | 0 | 0 | 0 | 0 | 0 | 0 | 0 | 39 | 471 | 0 | 510 | NavFac, 2018–2021 |
| Direct | Borehole | RHMW19 | 0 | 15 | 5 | 42 | 23 | 143 | 205 | 17 | 84 | 426 | 0 | 960 | NavFac, 2018–2021 |
| Direct | Borehole | OWD1 | 0 | 0 | 0 | 0 | 0 | 0 | 0 | 0 | ND | ND | 0 | 0 | NavFac, 2018–2021 |
| Direct | Borehole | OWD2A | 0 | 8 | 14 | 0 | 35 | 64 | 43 | 0 | ND | ND | 0 | 164 | NavFac, 2018–2021 |
| Direct | Borehole | OWD3A | 0 | 20 | 11 | 27 | 0 | 71 | 27 | 7 | ND | ND | 0 | 163 | NavFac, 2018–2021 |
| Direct | Borehole | OWD4 | 0 | 5 | 0 | 4 | 17 | 135 | 23 | 0 | ND | ND | 0 | 184 | NavFac, 2018–2021 |
| Direct | Borehole | OWD5 | 0 | 16 | 7 | 26 | 13 | 41 | 32 | 8 | ND | ND | 0 | 143 | NavFac, 2018–2021 |
| Direct | Borehole | OWD6 | 0 | 20 | 12 | 66 | 1 | 94 | 24 | 0 | ND | ND | 0 | 217 | NavFac, 2018–2021 |
| Direct | Borehole | OWD7 | 0 | 8 | 45 | 30 | 10 | 105 | 47 | 0 | ND | ND | 0 | 245 | NavFac, 2018–2021 |
| Direct | Borehole | OWD8 | 0 | 16 | 44 | 52 | 30 | 233 | 73 | 0 | ND | ND | 0 | 478 | NavFac, 2018–2021 |
| Indirect | Seismic refraction | Profile A | 0 | 908 | 4490 | 336 | 145 | 0 | 2963 | 84 | ND | ND | 0 | 8926 | Liberty and St. Clair, 2018 |
| Indirect | Seismic refraction | Profile B | 0 | 781 | 1032 | 44 | 41 | 0 | 1269 | 0 | ND | ND | 0 | 3167 | Liberty and St. Clair, 2019 |
| Indirect | Seismic refraction | Proifle C | 0 | 1135 | 1822 | 524 | 312 | 0 | 165 | 118 | ND | ND | 0 | 4076 | Liberty and St. Clair, 2020 |
| Indirect | Seismic refraction | Profile D | 0 | 1810 | 5046 | 560 | 79 | 0 | 0 | 0 | ND | ND | 0 | 7495 | Liberty and St. Clair, 2021 |
| Indirect | Seismic refraction | Profile E | 0 | 891 | 2766 | 0 | 0 | 0 | 0 | 0 | ND | ND | 0 | 3657 | Liberty and St. Clair, 2022 |
| Indirect | Seismic refraction | Profile F | 0 | 716 | 1684 | 366 | 49 | 0 | 0 | 2 | ND | ND | 0 | 2817 | Liberty and St. Clair, 2023 |
| Indirect | Seismic refraction | Profile G | 0 | 2172 | 1678 | 908 | 1420 | 0 | 1166 | 1009 | ND | ND | 0 | 8353 | Liberty and St. Clair, 2024 |
| Indirect | Seismic refraction | Profile H | 0 | 2515 | 2170 | 82 | 0 | 0 | 0 | 0 | ND | ND | 0 | 4767 | Liberty and St. Clair, 2025 |
| Indirect | Seismic refraction | Profile I | 0 | 650 | 1050 | 0 | 0 | 0 | 0 | 0 | ND | ND | 0 | 1700 | Liberty and St. Clair, 2026 |
| Indirect | Seismic reflection | saprolite-pahoehoe | 0 | 0 | 2994 | 0 | 0 | 0 | 2994 | 0 | ND | ND | 0 | 5988 | Liberty and St. Clair, 2027 |
| Indirect | Drilling | caprock-pahoehoe | 3692 | 0 | 0 | 0 | 0 | 0 | 776 | 1653 | ND | ND | 0 | 6121 | Hunt, 1997 |
| Indirect | Map | Alluvium | 0 | 94 | 0 | 0 | 0 | 0 | 0 | 0 | ND | ND | 0 | 94 | Sherrod et al., 2021 |
| Indirect | Map | Caprock | 110 | 0 | 0 | 0 | 0 | 0 | 0 | 0 | ND | ND | 0 | 110 | Sherrod et al., 2022 |
| Indirect | Map | Tuff | 0 | 0 | 0 | 0 | 0 | 0 | 0 | 90 | ND | ND | 0 | 90 | Sherrod et al., 2023 |
| Indirect | Harbor | Nodes | 0 | 0 | 0 | 0 | 0 | 0 | 0 | 0 | ND | ND | 28 | 28 | PacIOOS, 2023 |
| Indirect | Deep well | Kaamilo | 0 | 0 | 0 | 0 | 0 | 0 | 9 | 0 | ND | ND | 0 | 9 | Board of Water Supply, 2023 |
| Indirect | Deep well | Halawa | 0 | 0 | 0 | 0 | 0 | 0 | 9 | 0 | ND | ND | 0 | 9 | Board of Water Supply, 2023 |
| Indirect | Deep well | Moanalua | 0 | 0 | 0 | 0 | 0 | 0 | 9 | 0 | ND | ND | 0 | 9 | Board of Water Supply, 2023 |
| Indirect | Synthetic | Wells | 0 | 0 | 0 | 0 | 0 | 0 | 1152 | 0 | ND | ND | 0 | 1152 | This study |
| | | Sum = | 3802 | 12083 | 25310 | 3774 | 2507 | 2580 | 13295 | 3082 | 2737 | 6613 | 28 | 75811 | |
| | | Fraction = | 0.05 | 0.16 | 0.33 | 0.05 | 0.03 | 0.03 | 0.18 | 0.04 | 0.04 | 0.09 | 0.0004 | 1 | |

ND = no data.

**Table 3.**

Summary of numerical feature counts included in the Data cube. The table includes observation (direct or indirect), type (borehole, seismic refraction, gravity, contact depth, map, harbor, deep wells, and synthetic wells), name of observation, and reference.

| | | | Numerical Feature Count | | | | | | | | | | | |
|---|---|---|---|---|---|---|---|---|---|---|---|---|---|---|
| | | | Geophysiccal | > | < | Engineering | > | < | Water Quality | > | | | | |
| Observation | Type | Name | Velocity[1] | Densty[2] | Rock quality designation[3] | Fracture Rate[3] | Barometric pressure[4] | Temperature[4] | Conductance[4] | Dissolved oxygen[4] | Oxygen reduction potential[4] | pH[4] | Total | Reference |
| Indirect | Borehole | RHP01 | 177 | 3 | 154 | 157 | 0 | 25 | 24 | 3 | 1 | 1 | 545 | See list below. |
| Indirect | Borehole | RHP02 | 162 | 0 | 137 | 144 | 0 | 7 | 11 | 2 | 1 | 1 | 465 | See list below. |
| Indirect | Borehole | RHP03 | 157 | 0 | 141 | 141 | 0 | 14 | 1 | 1 | 1 | 1 | 457 | See list below. |
| Indirect | Borehole | RHP04A | 141 | 0 | 154 | 165 | 0 | 14 | 14 | 14 | 14 | 14 | 530 | See list below. |
| Indirect | Borehole | RHMW01R | 235 | 1 | 264 | 166 | 1 | 25 | 25 | 15 | 15 | 15 | 762 | See list below. |
| Indirect | Borehole | RHMW02 | 100 | 0 | 108 | 5 | 0 | 15 | 15 | 15 | 15 | 15 | 288 | See list below |
| Indirect | Borehole | RHMW03 | 84 | 0 | 116 | 20 | 0 | 13 | 13 | 13 | 13 | 13 | 285 | See list below |
| Indirect | Borehole | RHMW04 | 35 | 0 | 50 | 50 | 0 | 11 | 11 | 11 | 11 | 11 | 190 | See list below |
| Indirect | Borehole | RHMW05 | 51 | 0 | 102 | 5 | 0 | 18 | 18 | 18 | 18 | 18 | 248 | See list below |
| Indirect | Borehole | RHMW6 | 160 | 12 | 269 | 12 | 0 | 0 | 0 | 0 | 0 | 0 | 453 | See list below |
| Indirect | Borehole | RHMW7 | 39 | 5 | 1 | 1 | 0 | 0 | 0 | 0 | 0 | 0 | 46 | See list below |
| Indirect | Borehole | RHMW8 | 275 | 0 | 304 | 309 | 0 | 10 | 10 | 10 | 10 | 10 | 938 | See list below |
| Indirect | Borehole | RHMW09 | 405 | 10 | 413 | 411 | 0 | 11 | 11 | 11 | 11 | 11 | 1294 | See list below |
| Indirect | Borehole | RHMW10 | 489 | 9 | 495 | 499 | 0 | 10 | 10 | 10 | 10 | 10 | 1542 | See list below |
| Indirect | Borehole | RHMW11 | 425 | 193 | 488 | 459 | 0 | 0 | 0 | 0 | 0 | 0 | 1565 | See list below |
| Indirect | Borehole | RHMW12A | 431 | 1 | 446 | 442 | 0 | 11 | 11 | 11 | 11 | 11 | 1375 | See list below |
| Indirect | Borehole | RHMW13 | 509 | 22 | 529 | 22 | 0 | 0 | 0 | 0 | 0 | 0 | 1082 | See list below |
| Indirect | Borehole | RHMW14 | 422 | 27 | 473 | 41 | 0 | 0 | 0 | 0 | 0 | 0 | 963 | See list below |
| Indirect | Borehole | RHMW15 | 588 | 3 | 572 | 2 | 0 | 0 | 0 | 0 | 0 | 0 | 1165 | See list below |
| Indirect | Borehole | RHMW16 | 458 | 36 | 472 | 40 | 0 | 10 | 10 | 10 | 10 | 10 | 1056 | See list below |
| Indirect | Borehole | RHMW17 | 0 | 0 | 0 | 0 | 0 | 10 | 10 | 10 | 10 | 10 | 50 | See list below |
| Indirect | Borehole | RHMW19 | 435 | 5 | 445 | 15 | 0 | 11 | 11 | 11 | 11 | 11 | 955 | See list below |
| Indirect | Borehole | OWD1 | 0 | 0 | 0 | 0 | 0 | 0 | 0 | 0 | 0 | 0 | 0 | See list below |
| Indirect | Borehole | OWD2A | 152 | 0 | 159 | 15 | 0 | 0 | 0 | 0 | 0 | 0 | 326 | See list below |
| Indirect | Borehole | OWD3A | 136 | 0 | 143 | 0 | 0 | 0 | 0 | 0 | 0 | 0 | 279 | See list below |
| Indirect | Borehole | OWD4 | 179 | 0 | 169 | 0 | 0 | 0 | 0 | 0 | 0 | 0 | 348 | See list below |
| Indirect | Borehole | OWD5 | 121 | 0 | 126 | 0 | 0 | 0 | 0 | 0 | 0 | 0 | 247 | See list below |
| Indirect | Borehole | OWD6 | 188 | 0 | 22 | 0 | 0 | 0 | 0 | 0 | 0 | 0 | 210 | See list below |
| Indirect | Borehole | OWD7 | 237 | 0 | 244 | 0 | 0 | 0 | 0 | 0 | 0 | 0 | 481 | See list below |
| Indirect | Borehole | OWD8 | 221 | 1 | 84 | 1 | 1 | 1 | 1 | 1 | 1 | 1 | 313 | See list below |
| Indirect | Seismic refraction | Profile A | 7825 | 0 | 0 | 0 | 0 | 0 | 0 | 0 | 0 | 0 | 7825 | Liberty and St. Clair, 2018 |
| Indirect | Seismic refraction | Profile B | 2929 | 0 | 0 | 0 | 0 | 0 | 0 | 0 | 0 | 0 | 2929 | Liberty and St. Clair, 2019 |
| Indirect | Seismic refraction | Proifle C | 3955 | 0 | 0 | 0 | 0 | 0 | 0 | 0 | 0 | 0 | 3955 | Liberty and St. Clair, 2020 |
| Indirect | Seismic refraction | Profile D | 6893 | 0 | 0 | 0 | 0 | 0 | 0 | 0 | 0 | 0 | 6893 | Liberty and St. Clair, 2021 |
| Indirect | Seismic refraction | Profile E | 3193 | 0 | 0 | 0 | 0 | 0 | 0 | 0 | 0 | 0 | 3193 | Liberty and St. Clair, 2022 |
| Indirect | Seismic refraction | Profile F | 2443 | 0 | 0 | 0 | 0 | 0 | 0 | 0 | 0 | 0 | 2443 | Liberty and St. Clair, 2023 |
| Indirect | Seismic refraction | Profile G | 8353 | 0 | 0 | 0 | 0 | 0 | 0 | 0 | 0 | 0 | 8353 | Liberty and St. Clair, 2024 |
| Indirect | Seismic refraction | Profile H | 4771 | 0 | 0 | 0 | 0 | 0 | 0 | 0 | 0 | 0 | 4771 | Liberty and St. Clair, 2025 |
| Indirect | Seismic refraction | Profile I | 1700 | 0 | 0 | 0 | 0 | 0 | 0 | 0 | 0 | 0 | 1700 | Liberty and St. Clair, 2026 |
| Indirect | Gravity | Density model | 0 | 37027 | 0 | 0 | 0 | 0 | 0 | 0 | 0 | 0 | 37027 | Ito et al., 2027 |
| Indirect | Contact depth | Caprock-Pahoehoe | 6133 | 5203 | 0 | 0 | 0 | 0 | 0 | 0 | 0 | 0 | 11336 | Hunt, 1997 |
| Indirect | Map | Alluvium | 94 | 0 | 94 | 94 | 0 | 0 | 0 | 0 | 0 | 0 | 282 | Sherrod et al., 2021 |
| Indirect | Map | Caprock | 110 | 110 | 0 | 0 | 0 | 0 | 0 | 0 | 0 | 0 | 220 | Sherrod et al., 2022 |
| Indirect | Map | Tuff | 90 | 90 | 0 | 0 | 0 | 0 | 0 | 0 | 0 | 0 | 180 | Sherrod et al., 2023 |
| Direct | Deep well | Kaamilo profile | 0 | 0 | 0 | 0 | 4919 | 4919 | 4919 | 61 | 61 | 61 | 14940 | Board of Water Supply, 2023 |
| Direct | Deep well | Halawa profile | 0 | 0 | 9 | 0 | 4231 | 4231 | 4231 | 395 | 395 | 395 | 13887 | Board of Water Supply, 2023 |
| Direct | Deep well | Moanalua profile | 0 | 0 | 9 | 0 | 4741 | 4741 | 4741 | 309 | 309 | 309 | 15159 | Board of Water Supply, 2023 |
| Direct | Harbor | Seawater profile | 28 | 28 | 28 | 28 | 28 | 28 | 28 | 28 | 28 | 28 | 280 | PacIOOS, 2023 |
| | | Sum = | 55529 | 42786 | 7220 | 3244 | 13921 | 14135 | 14125 | 959 | 956 | 956 | 153831 | |
| | | Fraction = | 0.36 | 0.28 | 0.05 | 0.02 | 0.09 | 0.09 | 0.09 | 0.01 | 0.01 | 0.01 | 1 | |

References:  [1]Fathani et al., 2018; [2]Bücker et al., 1999; Miller et al., 1988; [3]NavFAC, 2018-2021; [4]Hunt, C.D., 2004

## 3.2 Reference model

### 3.2.1. Training and testing

Prior to training, continuous input features are normalized by their variance, and input feature vectors are randomly ordered to initialize the map weight vectors. The self-organizing map (SOM) is trained using a fixed number of neurons (map nodes) and predefined topological relationships. A toroidal grid topology—wrapping both vertically and horizontally—is employed, with hexagonally arranged neurons to preserve neighborhood continuity. The initial number of map units ($M$) is determined using the heuristic proposed by Vesanto and Alhoniemi (2000), where $M = 5\sqrt{N}$ and $N$ is the total number of training samples. This heuristic, which is based on the ratio of the two largest eigenvalues of the data covariance matrix, is used to define the initial grid dimensions (66 rows × 60 columns) for the Hālawa–Moanalua aquifer *reference model*.

To assess the influence of map resolution on training quality, two additional SOM grids are constructed and evaluated following the method of Friedel et al. (2020). The second grid is generated by scaling the initial grid dimensions by a factor of 1.5, and the third grid is generated by scaling the second grid by an additional factor of 1.5. These progressive increases in grid size result in corresponding increases in the total number of nodes (grid 1: 3,690 nodes; grid 2: 9,000 nodes; grid 3: 15,840 nodes) and in the computational cost required for model training and validation (Table 4).

**Table 4.**
Summary of unsupervised training topologies for each regional model. SOM=the self-organizing map.

| Map grid | Data cube | Initization | Grid shape | Lattice type | Map size | | |
|---|---|---|---|---|---|---|---|
| | | | | | Rows | Columns | Nodes |
| 1 | Hālawa-Moanalua | Random | Toroid | Hexagonal | 66 | 60 | 3960 |
| 2 | Hālawa-Moanalua | Random | Toroid | Hexagonal | 100 | 90 | 9000 |
| 3 | Hālawa-Moanalua | Random | Toroid | Hexagonal | 132 | 120 | 15840 |

Training of the self-organizing maps is conducted in two sequential phases: an initial rough training phase followed by a fine-tuning phase. A summary of the number of iterations for each phase, the corresponding Gaussian neighborhood radii, and the resulting quantization and topographic errors is provided in Table 5. For all map grids, learning rates decrease linearly from initial and final values of 0.5 and 0.05, respectively, and continue decaying to $10^{-5}$. In parallel, the Gaussian neighborhood function decreases exponentially (e.g., decay rate of $10^{-3}$ iteration$^{-1}$), yielding stable convergence across all network configurations.

**Table. 5.**
Summary of self-organizing map topologies for different size maps.

| Map grid | Data cube | Training phase | Neighborhood type | Radius initial | Radius final | Training length | Final quantization error | Final topograhic error |
|---|---|---|---|---|---|---|---|---|
| 1 | Hālawa-Moanalua | Rough | Gaussian | 90 | 23 | 20 | | |
| | Hālawa-Moanalua | Fine | Gaussian | 23 | 1 | 400 | 0.167 | 0.065 |
| 2 | Hālawa-Moanalua | Rough | Gaussian | 135 | 34 | 20 | | |
| | Hālawa-Moanalua | Fine | Gaussian | 34 | 1 | 400 | 0.109 | 0.062 |
| 3 | Hālawa-Moanalua | Rough | Gaussian | 179 | 45 | 20 | | |
| | Hālawa-Moanalua | Fine | Gaussian | 45 | 1 | 400 | 0.077 | 0.053 |

Evaluation during the final testing phase indicates that all three grid configurations converge to stable maps, as reflected by similar quantization and topographic error values. Summary statistics from unsupervised training across different network sizes produce comparable Cohen's kappa and overall accuracy values, indicating near-perfect predictability of geologic units (Cohen, 1960) regardless of the testing fold (Table 6). Given this consistency, the final model training is performed by presenting the complete set of Hālawa–Moanalua aquifer features (training and testing combined; 100%) to the smallest map grid (map grid 1). The resulting trained reference map is then used to simultaneously estimate transdisciplinary aquifer features—including geologic, geophysical, engineering, and water-quality attributes—referred to as the deterministic *reference model*.

**Table 6.**
Summary statistics for unsupervised training (modified self-organized map) applied to Hālawa-Moanalua aquifer features with different network sizes.

| Map grid | Selected Fold | Map size | Training n (80%) | Testing n (20%) | Kappa | Accuracy | Agreement |
|---|---|---|---|---|---|---|---|
| 1 | 10 | 66 x 60 | 51475 | 12866 | 0.84 | 0.91 | Almost perfect |
| 2 | 10 | 100 x 90 | 51475 | 12866 | 0.84 | 0.91 | Almost perfect |
| 3 | 10 | 132 x 120 | 51475 | 12866 | 0.87 | 0.93 | Almost perfect |
| 1 | 15 | 66 x 60 | 51475 | 12866 | 0.82 | 0.90 | Almost perfect |
| 2 | 15 | 100 x 90 | 51475 | 12866 | 0.85 | 0.92 | Almost perfect |
| 3 | 15 | 132 x 120 | 51475 | 12866 | 0.87 | 0.94 | Almost perfect |

*3.2.3. Grouping features*

A component-planes plot of the Hālawa–Moanalua (H–M) aquifer features is presented in Fig. 4. This plot provides a visual representation of nonlinear relationships among the multivariate input data at the local borehole scale and illustrate how features are organized across the self-organizing map. Patterns and associations among variables are examined by comparing the spatial distribution of values across the component planes (Vesanto and Alhoniemi, 2000). Component values are scaled such that dark red and dark blue represent the highest and lowest values of each input feature, respectively. This color scaling facilitates direct comparison of feature intensity and absence across the map.

For categorical variables—including geologic units (caprock, alluvium, saprolite, loose clinker, welded clinker, ʻaʻā, pāhoehoe, and tuff), aquifer confinement or saturation state (saturated or unsaturated), and ocean—the red color indicates the presence of a given category, while blue indicates its absence. During categorical training, geologic unit, saturation state, and ocean indicators are treated as mutually exclusive end members. As a result, the presence of any one category produces a red component plane, whereas its absence yields a blue plane. The spatial extent of red regions within each categorical component plane reflects the relative frequency of observations, with smaller regions indicating rarer features (e.g., ocean and ʻaʻā) and larger regions indicating more abundant units (e.g., pāhoehoe or saprolite). This representation provides an intuitive visualization of class imbalance within the H-M dataset.

Component planes are composed of networks of nodes arranged on a toroidal grid; therefore, the relative positioning of features across planes provides a means to evaluate whether the SOM has assimilated and grouped the input data appropriately. Consistent spatial alignment of features across multiple component planes indicates coherent organization within the trained network. For example, the localized red region associated with the welded clinker unit corresponds, at the same toroidal locations, to low-density values, unsaturated conditions, and low barometric pressure in other component planes. These systematic associations indicate that aquifer features are meaningfully separated and internally consistent across the map. Having confirmed appropriate feature assimilation and separation, principal component analysis is subsequently applied to the median values of the component planes to further evaluate dominant patterns in the data.

Principal component analysis (PCA) of the median component-plane values along the first two principal axes reveals relationships that are both internally consistent and physically intuitive (Fig. 5). The clear separation among aquifer feature vectors (shown as blue vectors) indicates that the associated hydrogeophysical observations contain sufficient and mutually informative structure to support development of a coherent Hālawa–Moanalua reference model. This separation suggests that the self-organizing map has successfully captured meaningful variability within the multivariate dataset. Moreover, the PCA results demonstrate that dominant patterns in the data align with expected hydrogeologic controls.

The PCA biplot further reveals four distinct groupings of aquifer features, with features within each group exhibiting stronger relationships to one another than to those in other groups. Features located in opposing groupings—for example, confined aquifer conditions (red group) versus unconfined aquifer conditions (brown group)—are antithetically, or negatively, correlated. Within each grouping, feature vectors that are closer together indicate stronger positive associations, whereas greater angular separation reflects weaker relationships. This structure provides a quantitative basis for interpreting how hydrogeologic properties co-vary across the aquifer system.

The confined aquifer grouping (red) indicates that increasing depth is associated with higher pāhoehoe abundance, greater rock quality designation (RQD), elevated barometric pressure, higher seismic velocity, increased density, and specific conductance. The increasing angular separation between density and conductance relative to the confined aquifer vector suggests that these relationships, while positive, are comparatively weaker. In contrast, the shallow unconfined aquifer grouping is characterized by strong associations among alluvium, saprolite, tuff, and temperature, reflecting near-surface conditions influenced by lithology and thermal gradients. Together, these groupings reinforce the physical consistency of the PCA results and their relevance to hydrogeologic interpretation.

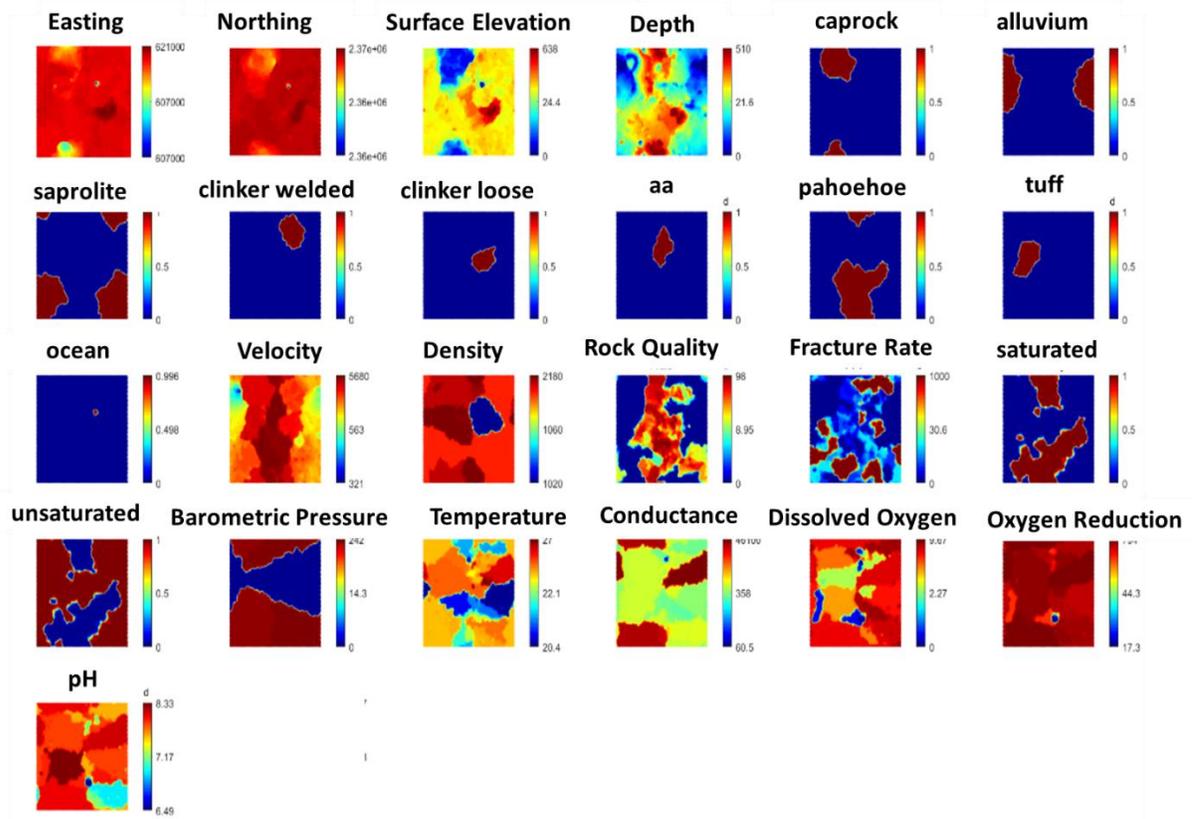

**Fig. 4.** Component planes of features across the SOM grid (toroid) characterizing the subregional Hālawa-Moanalua aquifer, Oahu, Hawai'i. Categorical features, e.g., geologic units ('a'ā, alluvium, caprock, clinker - loose, clinker - welded, pāhoehoe, saprolite, tuff), aquifer type (basal or unsaturated) and ocean (absent or present) and continuous features, e.g., easting, northing, surface elevation, depth, velocity, density, rock quality designation, fracture rate, barometric pressure, temperature, conductance, dissolved oxygen, oxygen reduction potential, and pH. Component planes are color coded where high values are hot (red) and low values are cool (blue). Same colors at same positions indicate positive correlation, and opposite colors at same positions indicate negative correlation. See Table 7 for feature units.

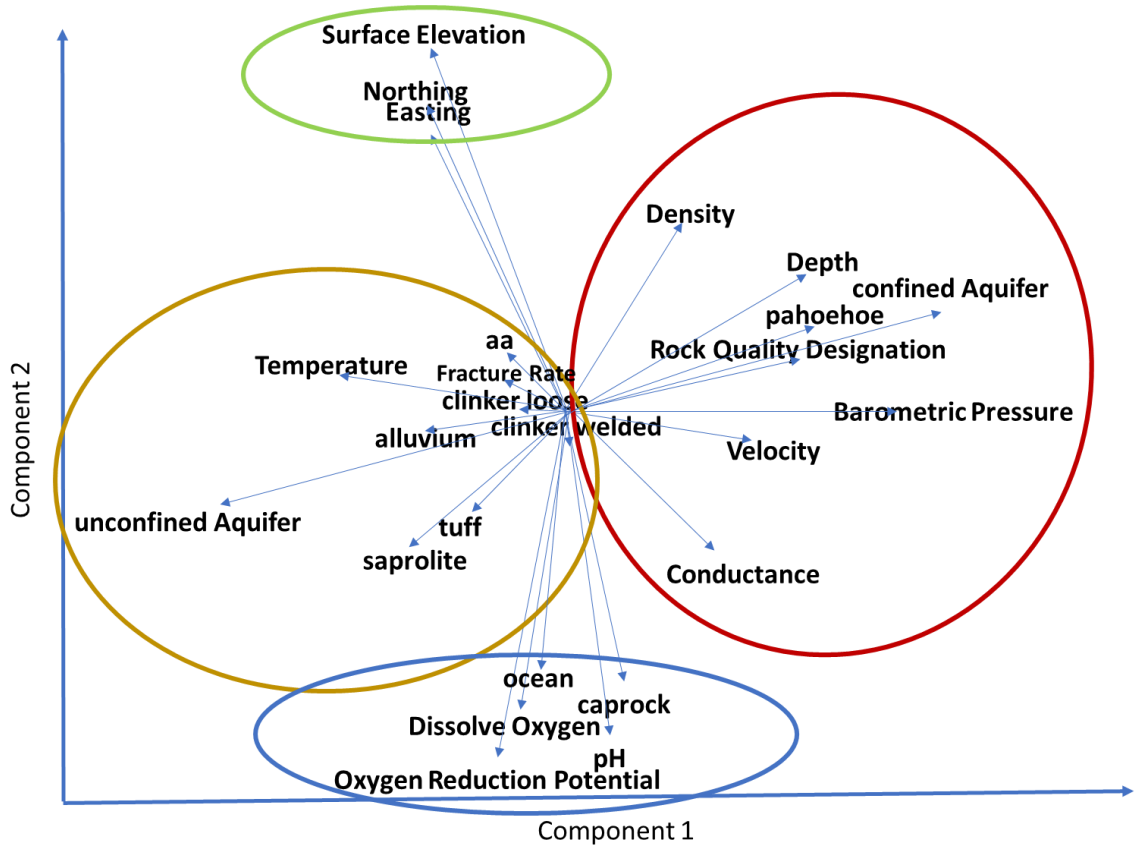

**Fig. 5.** Principal component analysis of features characterizing the subregional Hālawa-Moanalua aquifer. Categorical features (e.g., geologic units and aquifer confinement) are not capitalized, whereas continuous features are capitalized. Short feature vectors imply relatively few observations whereas long feature vectors reveal relatively many observations. The heterogeneity in vector lengths underscores the extremely imbalanced nature of observations in the data cube. Those features grouped together (e.g., grouped by same color circles) are more positively related whereas those on opposite sides are anti-correlated (e.g., brown and red circle, and green and blue circles). See Table 7 for feature units.

*3.3 Site model*

*3.3.1. Synthetic data generation methods*

A fixed number of synthetic observations are estimated using each AI generative algorithm, including the TGC, TVAE, CTGAN, and CopulaGAN. The synthetic datasets comprise both categorical features (e.g., geology and aquifer type) and numerical features (e.g., geophysical, engineering, and water-quality attributes) corresponding to the H-M reference model. For the Hālawa–Moanalua aquifer, nonparametric probability density functions and cumulative distribution functions are constructed from a fixed number of synthetic observations (N=1,000,000) for each feature (Figs. 6 and 7). In these analyses, *real observations* refer to measurements extracted from the data cube, whereas *synthetic observations* denote observations generated by the respective generative models. Metadata common to all four algorithms includes the enforcement of minimum and maximum feature values, no rounding, and 500 epochs.

The full set of aquifer features used in the analysis, including spatial, hydrogeologic, engineering, and water-quality attributes, along with their units and categorical encodings, is summarized in Table 7. By generating an equal number of synthetic observations for each model, distribution-based quality metrics are applied consistently to identify a preferred generative algorithm. The selected model is subsequently used to produce stochastic distributions from which a predetermined number of synthetic records are sampled and integrated with the reference model observations. This procedure is repeated while systematically increasing the number of synthetic point-cloud observations generated by each approach, enabling evaluation of algorithmic performance as a function of synthetic data volume.

**Table 7.**

Description of features used in the synthetic data generation and evaluation across the Hālawa–Moanalua aquifer.

| Feature | Description | Units / Encoding | Type |
| --- | --- | --- | --- |
| Easting | Easting coordinate | UTM | Numerical |
| Northing | Northing coordinate | UTM | Numerical |
| ELEVsurfm | Surface elevation | m | Numerical |
| DEPTHmbgs | Depth below ground surface | m | Numerical |
| VELOCITY | Seismic velocity | m/s | Numerical |
| DENSITY | Bulk density | kg/m$^3$ | Numerical |
| RQD | Rock Quality Designation | % | Numerical |
| FPF | Fracture rate | count/m | Numerical |
| ConfWLyes | Confined aquifer indicator | 1 = yes, 0 = no | Categorical |
| ConfWLno | Unconfined aquifer indicator | 1 = yes, 0 = no | Categorical |
| PRES | Barometric pressure | psi | Numerical |
| TEMPC | Temperature | °C | Numerical |
| COND | Specific conductance | µS/cm | Numerical |
| DO | Dissolved oxygen | mg/L | Numerical |
| ORP400 | Oxidation–reduction potential | mV | Numerical |
| PH | pH | – | Numerical |
| Geology_id | Geological unit | 0 = caprock; 1 = alluvium; 2 = saprolite; 3 = clinker–loose; 4 = clinker–welded; 5 = ʻaʻā; 6 = pāhoehoe; 7 = tuff | Categorical |

Geology is fundamental to groundwater-flow modeling because it controls groundwater pathways, flow velocities, and subsurface interactions. Accordingly, the reliability of a groundwater model depends on the accuracy of its geologic representation. To assess the performance of generative algorithms in reproducing observed geologic variability, bar charts comparing the proportional distributions of observed (real) and estimated (fake) geologic units are presented (Fig. 6). The results indicate that CopulaGAN most accurately reproduces geologic unit proportions under conditions of extreme class imbalance. TGC and TVAE also provide reasonable approximations; however, TGC overestimates the proportion of caprock, the most frequently observed unit, while TVAE overestimates pahoehoe, the second most abundant unit. In contrast, CTGAN exhibits an approximately linear response across categories, leading to underrepresentation of units with large sample sizes and overrepresentation of those with small sample sizes.

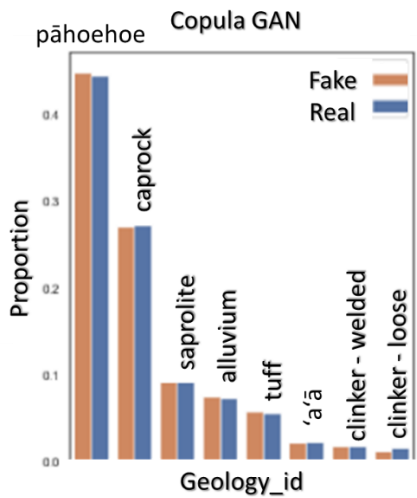
(a)

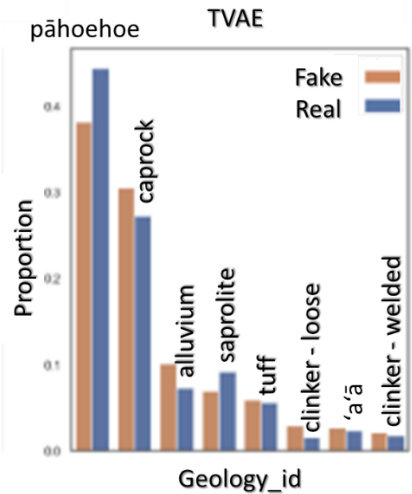
(b)

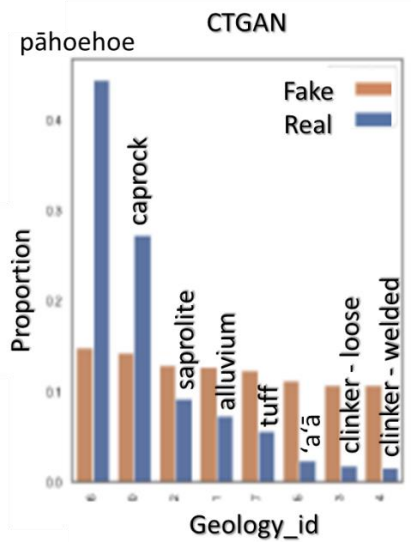
(c)

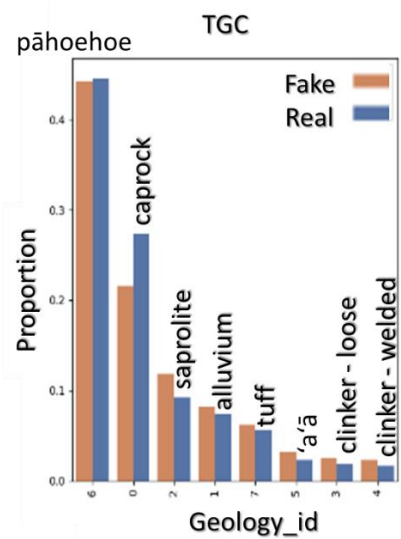
(d)

**Fig. 6.** Comparison of Real and Fake geologic unit proportions. (a) Copula Generative Adversarial Network - Copula GAN, (b) Tabular Variational Autoencoder - TVAE, (c) Constrained Generative Adversarial Network - CTGAN, and (d) Tabular Gaussian Copula - TGC. *Real* denotes observed observations extracted from the data cube, whereas *Fake* denotes synthetic observations generated by the respective algorithms. The order of each bar chart is set according to proportion of fake geologic units.

    Inspection of the nonparametric probability density functions (Fig. 7) and cumulative distribution functions (Fig. 8) reveals clear differences among the generative algorithms in their ability to reproduce observed (real) distributions for both categorical and numerical features. For the categorical representation of Hālawa–Moanalua geologic units (Geology_id), model performance ranks, from highest to lowest, as CopulaGAN, TVAE, Gaussian Copula, and CTGAN. For numerical variables, the deep generative models generally capture the multimodal structure of the observed distributions, although reproduction of extreme values is less consistent (Fig. 7b–d). In contrast, the TGC model shows limited capacity to reproduce multimodality in the continuous aquifer features, consistent with its assumption of an underlying Gaussian dependence structure (Fig. 7b–d). These observations are further supported by inspecting the cumulative distribution functions for both numerical and categorical AI-generated features (Fig. 8a–d). Given the importance of geologic heterogeneity in groundwater flow and transport simulations, and the comparatively weaker performance of TGC in representing key continuous features, CopulaGAN is selected as the preferred feature generative approach for subsequent analyses. Quantitative performance metrics are then applied to formally evaluate and corroborate the visually inferred differences among models.

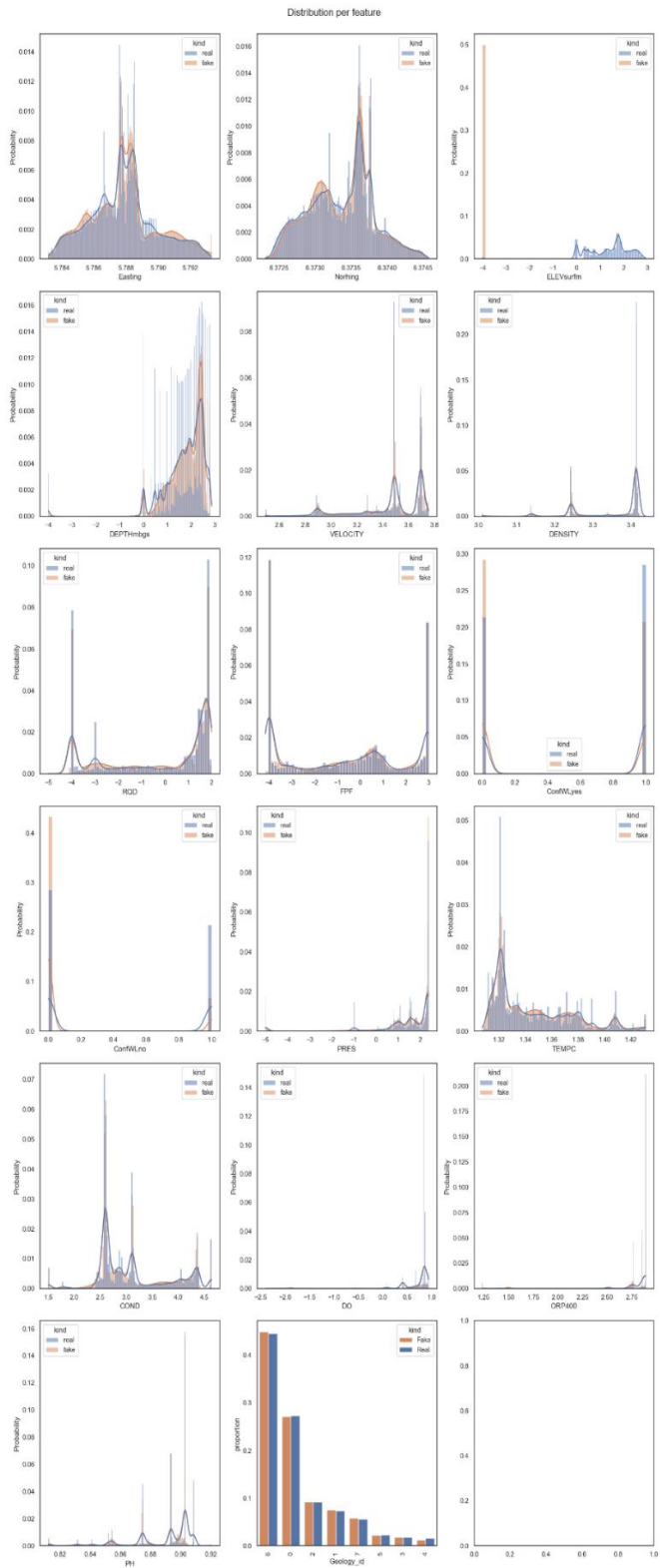

(a) Copula Generative Adversarial Network - Copula GAN

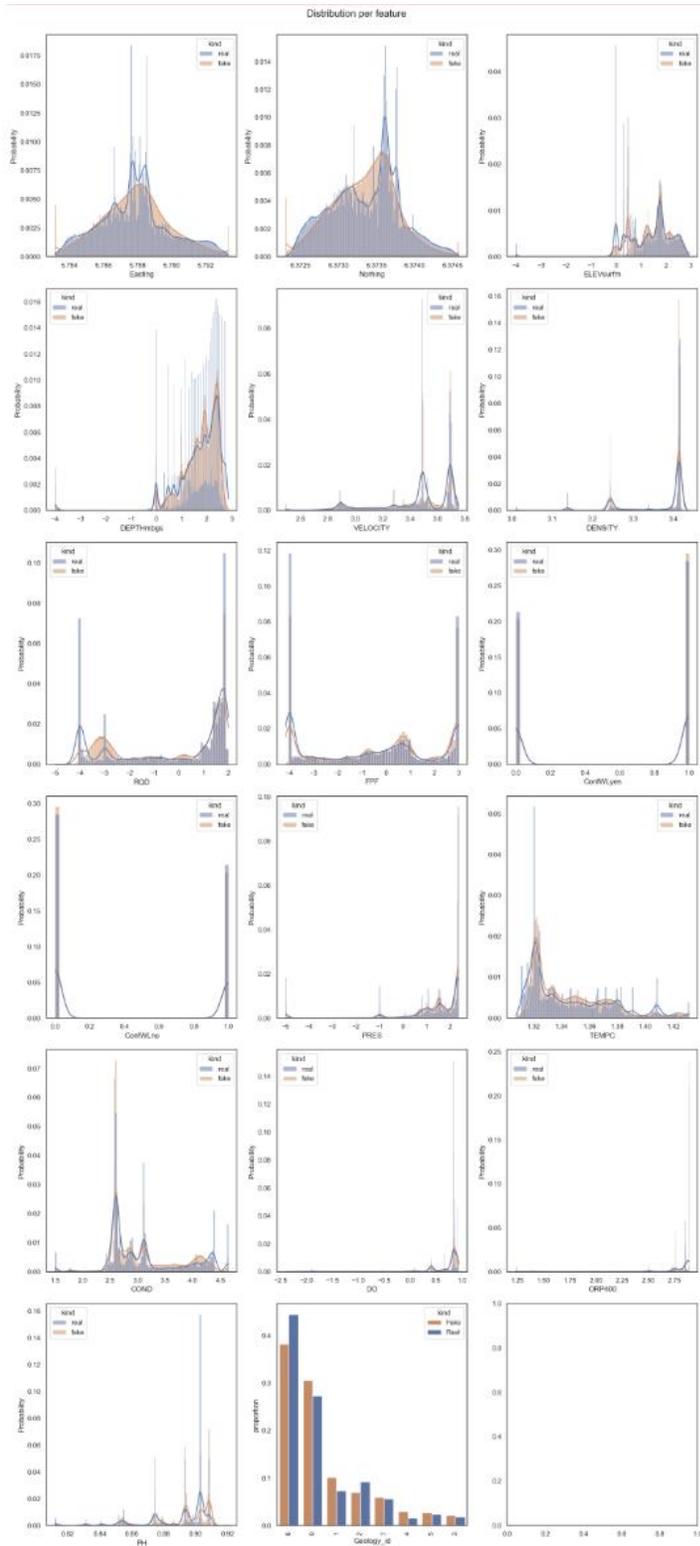

(b) Tabular Variational Autoencoder – TVAE

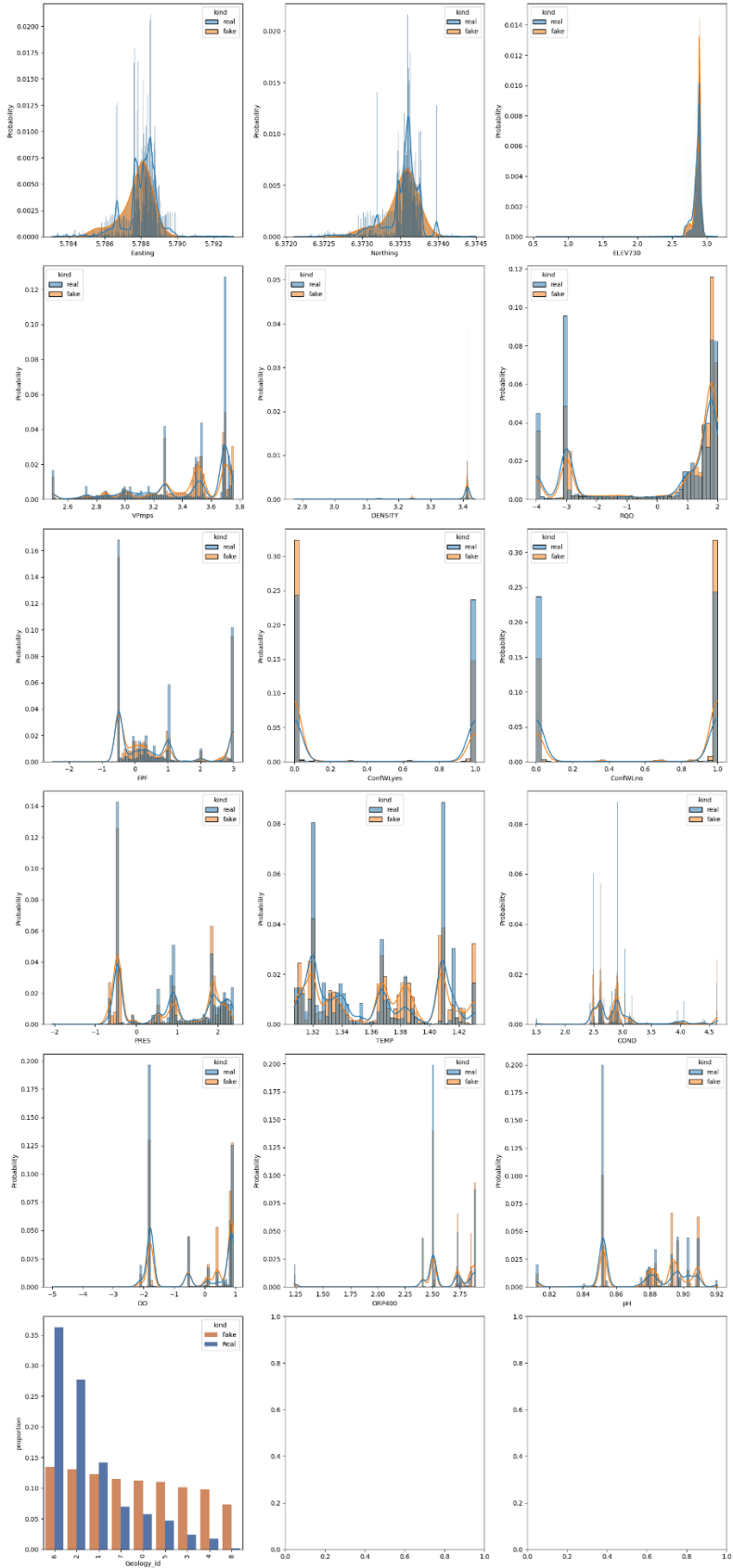

(c) Constrained Tabular Generative Adversarial Network - CTGAN

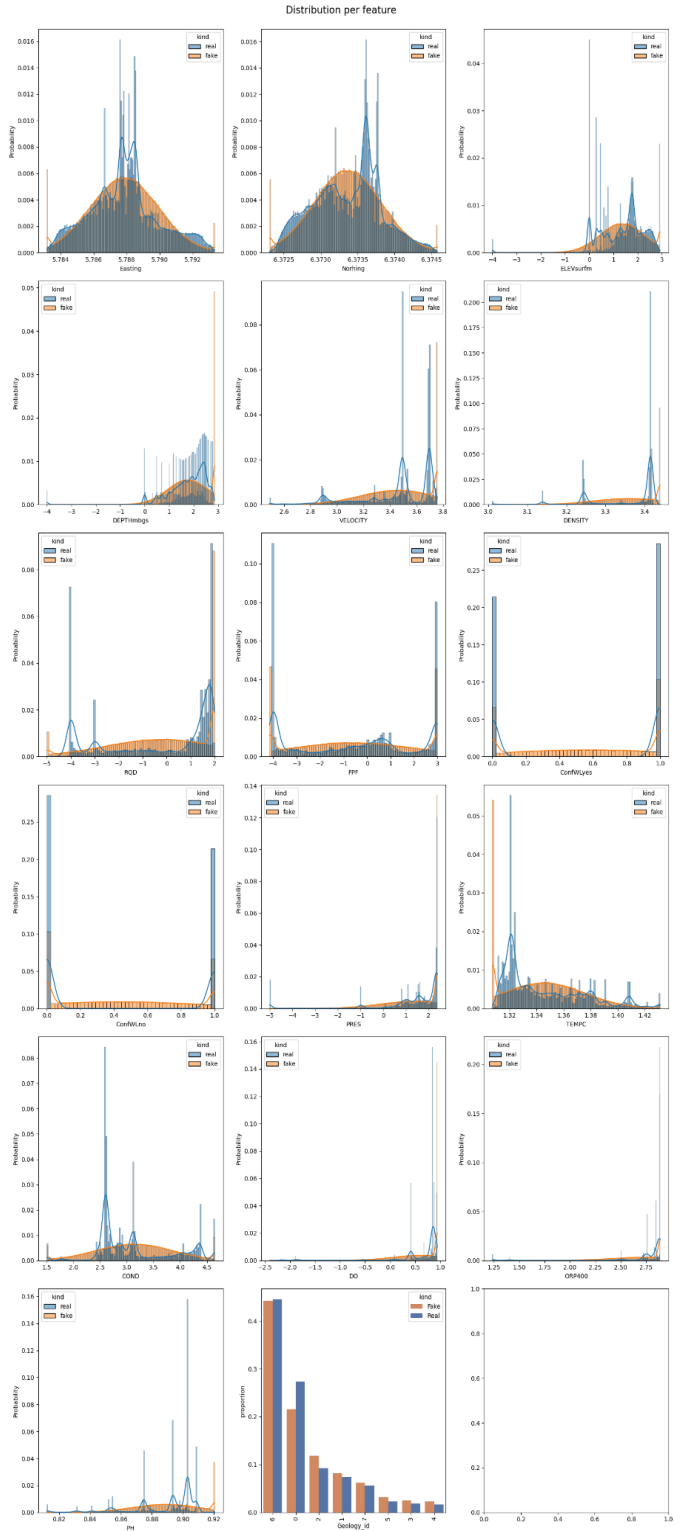

(d) Tabular Gaussian copula - TGC.

**Fig. 7.** Distribution functions of numerical and categorical features generated by the generative AI models (N = 1,000,000) for the Hālawa–Moanalua aquifer. *Real* denotes observed data extracted from the data cube, whereas *Fake* denotes synthetic observations generated by the respective algorithms. Variables include spatial coordinates, subsurface and engineering properties, hydrogeologic conditions, water-quality parameters, and geological unit classifications (see Table 7 for feature descriptions, units, and categorical encodings). (a) Copula Generative Adversarial Network - CopulaGAN, (b) Tabular Variational Autoencoder - TVAE, (c) Constrained Tabular Generative Adversarial Network – CTGAN; (d) Tabular Gaussian copula – TGC.

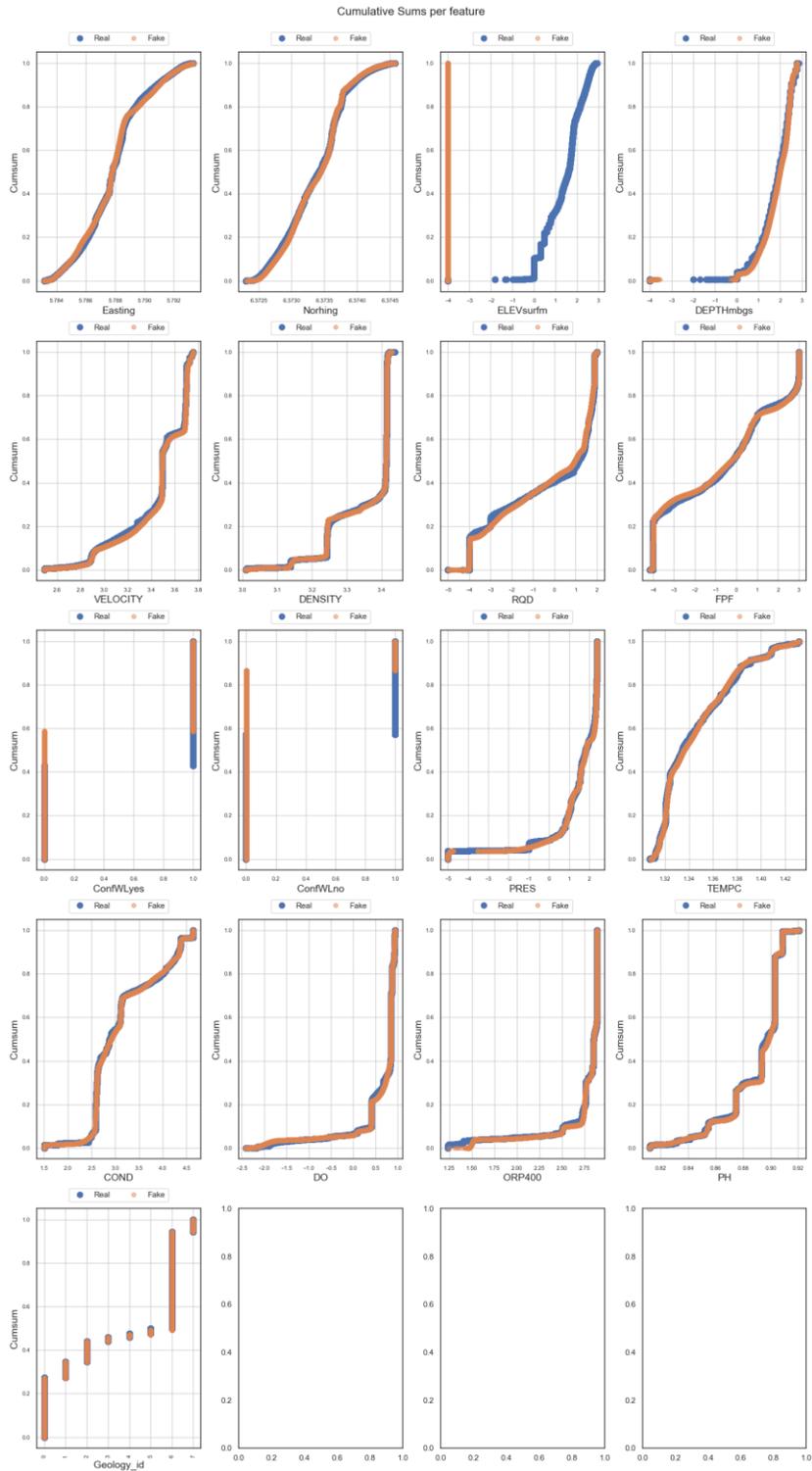

(a) Copula Generative Adversarial Network - Copula GAN

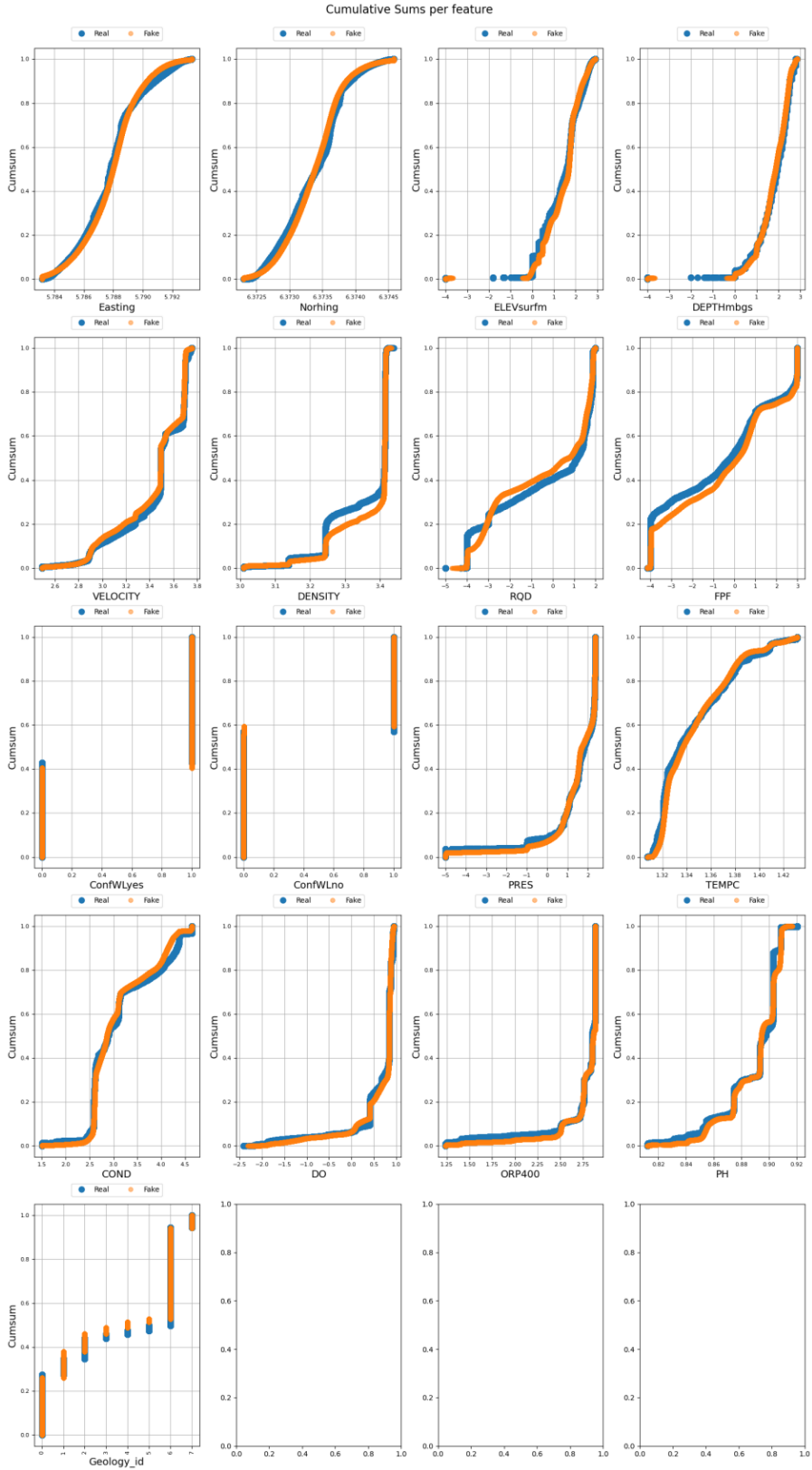

(a) Tabular Variational Autoencoder - TVAE

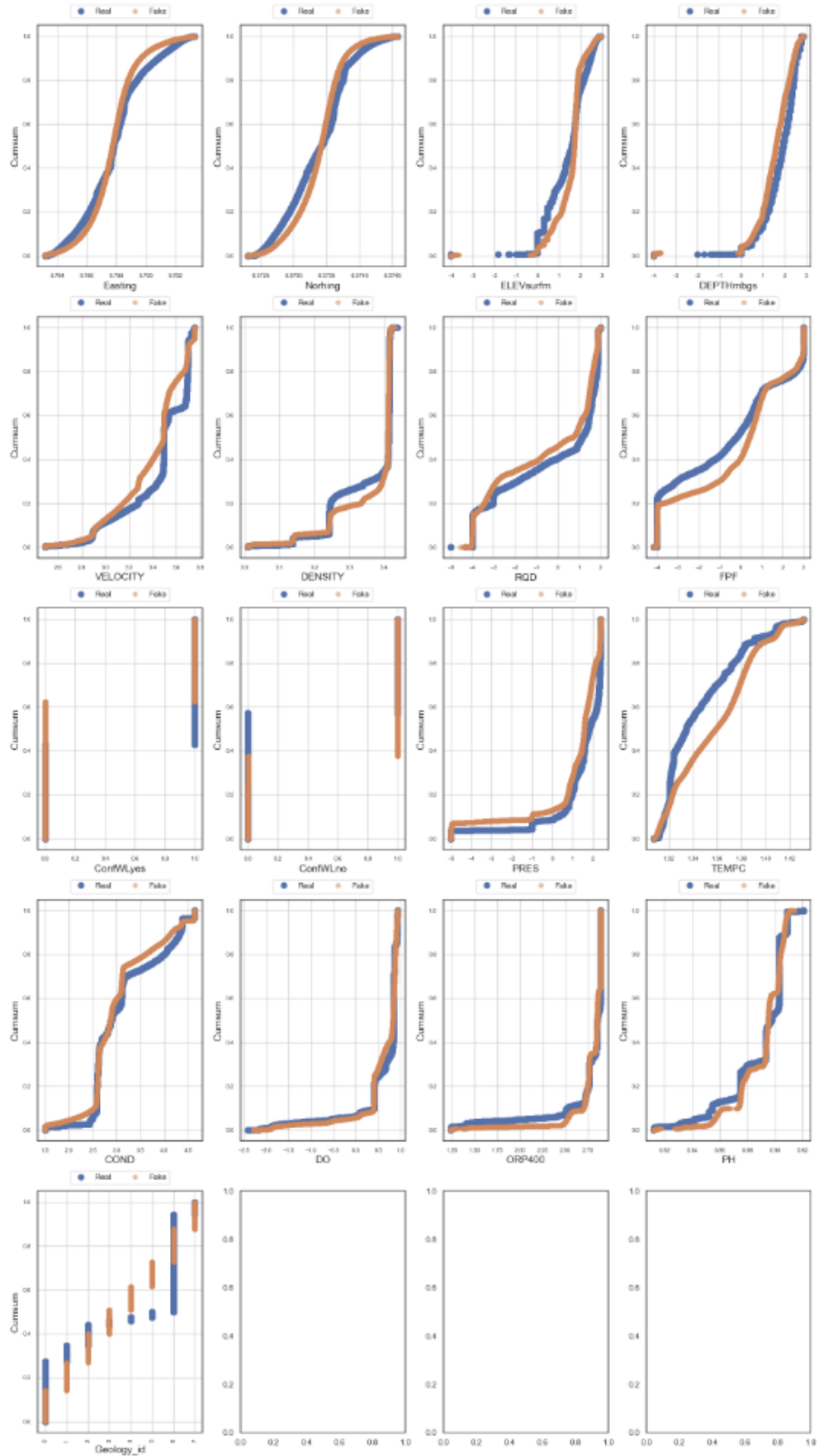

(b) Constrained Tabular Generative Adversarial Network - CTGAN

Cumulative Sums per feature

(c) Tabular Gaussian copula - TGC.

**Fig. 8.** Cumulative distribution functions of AI generated (N=1,000,000) numerical and categorical features for the Hālawa-Moanalua aquifer. *Real* denotes observed data extracted from the data cube, whereas *Fake* denotes synthetic observations generated by the respective algorithms. Variables include spatial coordinates, subsurface and engineering properties, hydrogeologic conditions, water-quality parameters, and geological unit classifications (see Table 7 for feature descriptions, units, and categorical encodings). (a) Copula Generative Adversarial Network - CopulaGAN, (b) Tabular Variational Autoencoder - TVAE, (c) Constrained Tabular Generative Adversarial Network – CTGAN; (d) Tabular Gaussian copula – TGC.

*3.3.2. Validation of tabular generative algorithms*

This section presents results from the application of multiple generative algorithms to the reference model. The results are summarized using tables, cumulative distribution functions, and probability distribution functions. Model performance is evaluated using the column shape score, column pair trends score, and average column score, based on AI-generated ensembles of 1,000,000 synthetic point-cloud values for both categorical and numerical features across the Hālawa–Moanalua aquifer (Table 8). At this scale of analysis, all generative algorithms exhibit broadly comparable performance, with TVAE showing marginally improved scores across the evaluated metrics.

Validation metrics for the synthetic AI generation of point-cloud values representing aquifer characteristics of the Hālawa–Moanalua aquifer are summarized in Table 9. In this analysis, the Kolmogorov–Smirnov (KS) complement metric is used to assess the ability of the generative algorithms to reproduce continuous aquifer properties and categorical aquifer confinement. KS complement values range from 0.66 to 0.96, where 0 indicates no correspondence and 1.0 indicates perfect correspondence; italicized values denote the lowest agreement, and bolded values denote the highest agreement. At this finer level of aquifer characterization, CopulaGAN performs best for the generation of numerical features, followed by TVAE, CTGAN, and TGC. In contrast, for categorical confinement, TVAE shows the strongest performance, followed by CTGAN, TGC, and CopulaGAN. These results indicate that different generative algorithms may be preferable for synthesizing numerical versus categorical aquifer features.

Summary statistics are used to compare numerical aquifer feature values estimated by the two-phase SOM-based deterministic reference model and the CopulaGAN-based stochastic site model (Table 10). Univariate statistics are computed for reference model values at field locations (N = 348,876) and for two CopulaGAN realizations with different numbers of synthetic point-cloud estimates (N = 525,000 and N = 695,000). Statistical measures, including sample size, mean, standard deviation, minimum, quartiles, and maximum, show strong agreement between the reference model and CopulaGAN estimates, indicating high repeatability of the stochastic generation process. The two CopulaGAN realizations exhibit nearly identical statistical characteristics, whereas larger differences between deterministic and stochastic estimates are attributed to uncertainty inherent in the reference model.

**Table 8.**

Comparison of metrics for AI generation of 1000000 synthetic observations of point cloud values for categorical and continuous features across the sub-regional Hālawa-Moanalua aquifer, Oahu, Hawai'i.

| Geneartive AI algorithm | Column shapes score (%) | Column pair trends (%) | Average score (%) | Count |
|---|---|---|---|---|
| Tabular Gaussian Copula | 79.0 | 93.3 | 86.2 | 1,000,000 |
| Tabular Variational Autoencoder | 86.6 | 97.2 | 91.4 | 1,000,000 |
| Constrained Generative Adversarial Network | 79.1 | 88.5 | 83.8 | 1,000,000 |
| Copula Generative Adversarial Network | 79.6 | 96.9 | 88.4 | 1,000,000 |

**Table 9.**

Validation metrics for synthetic AI generation (1000000 observations) of aquifer features at the Hālawa-Moanalua aquifer. The KS complement metrics range from 0.66 to 0.96 (1.0=perfect) with red values revealing lowest and black bolded revealing highest values. From this table, the order of best to worst algorithm for synthetic generation of continuous feature values are the CopulaGAN (preferred), TVAE, CTGAN and Gaussian copula.

| Type | Feature | Synthetic observations: | 400k | 400k | 400k | 400k | 500k | 1000k |
|---|---|---|---|---|---|---|---|---|
| | | Algorithm: | Gaussian copula | TVAE | CTGAN | CoplulaGAN | CoplulaGAN | CoplulaGAN |
| | | | KS complement (%) | KS complement (%) | KS complement (%) | KS complement (%) | KS complement (%) | KS complement (%) |
| Numerical | Easting (UTM) | | 0.93 | 0.95 | 0.92 | **0.96** | **0.96** | **0.96** |
| | Northing (UTM) | | 0.92 | 0.91 | 0.87 | **0.98** | **0.98** | **0.98** |
| | Surface elevation (m) | | 0.90 | **0.91** | 0.86 | - | - | - |
| | Depth (m) | | 0.89 | 0.93 | 0.81 | **0.94** | **0.94** | **0.94** |
| | Velocity (m/s) | | 0.79 | **0.92** | 0.84 | 0.91 | 0.91 | 0.91 |
| | Density (kg/m3) | | 0.66 | 0.91 | 0.75 | **0.92** | **0.92** | **0.92** |
| | Rock quallity designation | | 0.76 | **0.88** | 0.87 | 0.89 | 0.89 | 0.89 |
| | Fracture rate (count/m) | | 0.86 | 0.86 | 0.87 | **0.89** | **0.89** | **0.89** |
| | Barometric pressure (psi) | | 0.74 | 0.89 | 0.78 | **0.94** | **0.94** | **0.94** |
| | Temperature (C) | | 0.85 | 0.86 | 0.80 | **0.96** | **0.96** | 0.95 |
| | Conductance | | 0.82 | 0.92 | 0.92 | **0.95** | **0.95** | **0.95** |
| | Dissolved Oxygen | | 0.72 | **0.86** | 0.85 | 0.85 | 0.85 | 0.85 |
| | pH | | 0.74 | 0.78 | **0.86** | 0.84 | 0.84 | 0.84 |
| | Oxygen reduction potential | | 0.67 | **0.79** | 0.71 | 0.71 | 0.71 | 0.71 |
| Categorical | Confined aquifer | | 0.62 | **0.74** | 0.55 | 0.57 | 0.57 | 0.57 |
| | Unconfined aquifer | | 0.62 | **0.74** | 0.68 | 0.43 | 0.43 | 0.43 |

**Table 10.**

Statistical comparison of reference model to site model statistics for Hālawa-Moanalua aquifer features. The reference model reflects deterministic model values following application of the two phase self-organizing map to H-M observations from the data cube. The Site model reflects application of the Copula Generative Adversarial Network (CopulaGAN) algorithm to the reference model. The slight differences among the two synthesized CouplaGAN models (realizations) in prediction statistics are associated with uncertainty in the deterministic reference model.

| Model | Algorithm | | Easting | Northing | Depth (m) | Velocity (m/s) | Density (kg/m3) | Rock quality desgination | Fracture rate (#/m) | Barometric pressure (psi) | Temperature (C) | Conductance (uS/cm) | Dissolved oxygen (mg/l) | Oxygen reduction potential (mv) | pH |
|---|---|---|---|---|---|---|---|---|---|---|---|---|---|---|---|
| Reference | Self organizing map | count | 348876 | 348876 | 348876 | 348876 | 348876 | 348876 | 348876 | 348876 | 348876 | 348876 | 348876 | 348876 | 348876 |
| | | mean | 613562.9 | 2362470.3 | 58 | 2844 | 2288 | 1 | 0 | 25.5 | 22.2 | 1361 | 4.19 | 164.1 | 7.74 |
| | | Stdev | 1.0 | 1.0 | 7 | 2 | 1 | 207 | 417 | 33.2 | 1.1 | 5 | 3.64 | -397.9 | 1.05 |
| | | Minimum | 606965.3 | 2356715.9 | 0 | 310 | 1024 | 0 | 0 | 0.0 | 20.3 | 32 | 0.00 | -382.8 | 6.49 |
| | | 25% | 611671.8 | 2360596.0 | 25 | 2300 | 1923 | 0 | 0 | 12.2 | 20.9 | 400 | 3.41 | 179.0 | 7.49 |
| | | 50% | 613509.0 | 2362750.0 | 90 | 3100 | 2590 | 14 | 1 | 65.5 | 21.6 | 766 | 7.00 | 312.2 | 7.89 |
| | | 75% | 614951.6 | 2364150.7 | 225 | 4904 | 2600 | 58 | 47 | 221.4 | 23.2 | 4617 | 7.50 | 375.0 | 8.00 |
| | | Maximum | 621336.6 | 2369077.0 | 732 | 5700 | 2748 | 100 | 1000 | 241.7 | 27.0 | 44000 | 8.78 | 375.0 | 8.33 |
| Site | Copula Generative Adversarial Network | count | 525000 | 525000 | 525000 | 525000 | 525000 | 525000 | 525000 | 525000 | 525000 | 525000 | 525000 | 525000 | 525000 |
| | | mean | 613554.2 | 2362506.8 | 67 | 2922 | 2289 | 1 | 0 | 27.1 | 22.2 | 1384 | 4.21 | 179.1 | 7.75 |
| | | Stdev | 1.0 | 1.0 | 6 | 2 | 1 | 174 | 455 | 31.4 | 1.1 | 5 | 3.83 | -398.0 | 1.05 |
| | | Minimum | 606965.3 | 2356715.9 | 0 | 310 | 1024 | 0 | 0 | 0.0 | 20.3 | 32 | 0.00 | -382.8 | 6.49 |
| | | 25% | 611557.6 | 2360682.9 | 29 | 2437 | 1941 | 0 | 0 | 11.9 | 21.0 | 403 | 4.17 | 178.1 | 7.50 |
| | | 50% | 613457.9 | 2362881.5 | 98 | 3105 | 2587 | 9 | 1 | 68.6 | 21.7 | 823 | 7.00 | 312.2 | 7.92 |
| | | 75% | 614813.6 | 2364085.5 | 243 | 4934 | 2600 | 53 | 101 | 222.5 | 23.2 | 4657 | 7.49 | 375.0 | 8.00 |
| | | Maximum | 621336.6 | 2369077.0 | 612 | 5700 | 2694 | 100 | 1000 | 241.6 | 27.0 | 44000 | 8.77 | 375.0 | 8.33 |
| Site | Copula Generative Adversarial Network | count | 695000 | 695000 | 695000 | 695000 | 695000 | 695000 | 695000 | 695000 | 695000 | 695000 | 695000 | 695000 | 695000 |
| | | mean | 613554.3 | 2362507.4 | 68 | 2922 | 2290 | 1 | 0 | 27.2 | 22.2 | 1383 | 4.21 | 178.5 | 7.75 |
| | | Stdev | 1.0 | 1.0 | 6 | 2 | 1 | 174 | 454 | 31.4 | 1.1 | 5 | 3.83 | -398.0 | 1.05 |
| | | Minimum | 606965.3 | 2356715.9 | 0 | 310 | 1024 | 0 | 0 | 0.0 | 20.3 | 32 | 0.00 | -381.6 | 6.49 |
| | | 25% | 611567.5 | 2360681.6 | 29 | 2432 | 1938 | 0 | 0 | 12.0 | 21.0 | 403 | 4.19 | 177.9 | 7.50 |
| | | 50% | 613457.2 | 2362883.4 | 98 | 3105 | 2587 | 9 | 1 | 69.0 | 21.7 | 820 | 7.00 | 312.2 | 7.92 |
| | | 75% | 614808.7 | 2364085.2 | 243 | 4934 | 2600 | 53 | 102 | 222.5 | 23.2 | 4649 | 7.49 | 375.0 | 8.00 |
| | | Maximum | 621336.6 | 2369077.0 | 611 | 5700 | 2695 | 100 | 1000 | 241.6 | 27.0 | 44000 | 8.77 | 375.0 | 8.33 |

*3.4 Conceptual groundwater model*

Based on a review of the validation metrics and summary statistical measures provided in the previous section, the preferred generative AI algorithm for the H-M aquifer is determined to be the CopulaGAN. Although the TVAE is preferred for estimating the confinement or unconfinement of the H-M aquifer. For this reason, the CopulaGAN is used to generate the synthetic set of stochastic H-M aquifer features as point clouds that are mapped to structured grid and called the Conceptual groundwater model. The one exception is for aquifer confinement. In this case, the TVAE is used to generate the synthetic set of stochastic H-M values for aquifer confinement that are mapped separately to the grid.

*3.4.1. Mapping synthetic point clouds to the structured model grid*

The final step in the AI-assisted workflow involves mapping the set of stochastic *site model* point clouds to nodes of the structured numerical groundwater model grid. The mapping of *site model* point clouds to grid nodes is performed using the two-phase SOM algorithm, resulting in a transdisciplinary representation of continuous and categorical aquifer features at each grid node. The development of a stochastic conceptual groundwater model on the structured grid involves the following workflow.

1. Site model realization**:** A site model realization is generated by applying the CopulaGAN to the reference model, producing synthetic point clouds of aquifer features.
2. Conceptual groundwater model realization**:** A self-organizing map (SOM) whose default grid size is applied to the site model realization to map the synthetic point clouds for each aquifer feature onto the nodes of the structured groundwater model grid.
3. Ensemble generation**:** Steps 1 and 2 are repeated to generate a statistically representative ensemble of conceptual groundwater models (N = 30).

4. **Stochastic model construction:** Stochastic conceptual groundwater models are produced by aggregating the ensemble results into cumulative distribution functions (CDFs) for each categorical and continuous aquifer feature at every grid node.
5. **Feature selection and visualization:** Aquifer features are extracted from their CDFs at selected percentiles for visualization, evaluation, and interpretation of individual three-dimensional categorical or continuous properties.

Each synthetic point cloud realization is mapped to the nodes of the structured groundwater model grid (N=310,500). Predicted categorical variables range from 0 to 1 (Fig. 9). Where predictions indicate multiple geologic units—reflecting uncertainty due to limited mutual information—the unit with a frequency greater than or equal to 0.5 is assigned, consistent with one-hot encoding. Empirical distribution functions for each aquifer feature are derived from the ensemble of realizations. The resulting CopulaGAN-based predictions provide spatially distributed stochastic representations of geologic units within and across the Hālawa–Moanalua aquifer. Percentile mapped voxels represent feature values extracted from the corresponding cumulative distribution functions computed from the realizations.

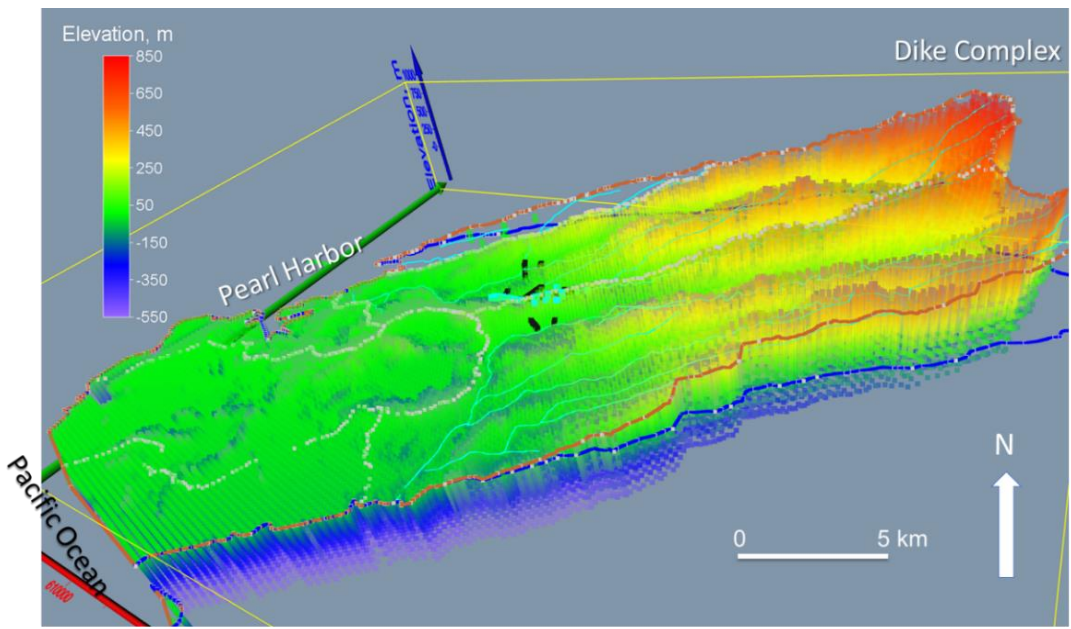

**Fig. 9.** The structured numerical model grid is constructed with 30 vertical layers and a horizontal resolution of 10 m × 10 m, resulting in a total of 310,500 nodes. Grid elevations are represented using a color scale, with warmer colors indicating higher elevations and cooler colors indicating lower elevations. The approximate locations of subregional features, including the dike complex, Pearl Harbor, and the Pacific Ocean, are also indicated.

The quality of geologic unit mapping from CopulaGAN-based point clouds to the structured Hālawa-Moanalua aquifer grid nodes are summarized in Table 11. The summary information includes the number of classified predictions by geologic unit (y-axis) as a function of deciles (x-axis). Inspection of this table reveals that the number of classified grid nodes for each geologic unit increases from the 10th percentile (minimum number) to the 90th percentile (maximum number). This behavior is characteristic of the cumulative distribution function meaning that the number of classified geologic units will always decrease when selecting lessor percentiles. For example, the estimated number of grid nodes classified as pāhoehoe is 133,608 at the 10th percentile and 191,091 at the 90th percentile with a range of prediction uncertainty of 57,483 nodal classifications (difference between 90th and 10th percentiles).

Analysis of the total number of structured grid nodes and the number of classified geologic units at selected percentiles across all nodes provides useful context for assessing the point-cloud–to-grid mapping process. At the 10th, 50th, and 90th percentiles, the total number of classified (unclassified) nodes are 211,378 (99,122 or 32%), 274,918 (35,582 or 11%), and 358,753 (−48,253 or −16%), respectively. The difference in the absolute percentage of unclassified nodes between the 10th and 90th percentiles indicates a slight bias toward a preferred set of feature predictions above the 75th percentile. Negative unclassified values beyond the 75th percentile suggest that multiple classifications are being assigned to one or more grid nodes. This behavior indicates the need to reevaluate in future studies: (1) the one-hot encoding assignment for geologic unit frequency greater than (or equal) to 0.5, and (2) the SOM grid size and processing parameters as the number of generative estimates increases.

**Table 11.**
CopulaGAN-based stochastic predictions of geologic units across the Hālawa–Moanalua aquifer. Percentiles of geologic units reflect cumulative distributions functions computed from 30 realizations.

| | Percentile | | | | | | | | | |
|---|---|---|---|---|---|---|---|---|---|---|
| Geologic Unit | 10th | 20th | 30th | 40th | 50th | 60th | 70th | 80th | 90th | Average |
| Caprock | 76540 | 84472 | 89159 | 93093 | 96500 | 99872 | 103603 | 108079 | 113904 | 96189 |
| Saprolite | 548 | 1024 | 1716 | 2345 | 3213 | 4228 | 5577 | 7963 | 12254 | 3081 |
| Clinker Loose | 0 | 0 | 0 | 0 | 6 | 150 | 374 | 1873 | 7956 | 6 |
| Clinker Welded | 1 | 319 | 667 | 1046 | 1457 | 1921 | 2507 | 3398 | 4885 | 1413 |
| ʻAʻā | 0 | 19 | 170 | 399 | 797 | 1331 | 2169 | 3992 | 6462 | 756 |
| Pāhoehoe | 133608 | 144598 | 154558 | 161928 | 166594 | 170810 | 175382 | 181056 | 191091 | 166399 |
| Tuff | 681 | 1922 | 3678 | 4947 | 6351 | 8088 | 10683 | 15070 | 22201 | 6246 |
| Total classified | 211378 | 232354 | 249948 | 263758 | 274918 | 286400 | 300295 | 321431 | 358753 | 274090 |
| Total nodes | 310500 | 310500 | 310500 | 310500 | 310500 | 310500 | 310500 | 310500 | 310500 | 310500 |
| Unclassified nodes | 99122 | 78146 | 60552 | 46742 | 35582 | 24100 | 10205 | -10931 | -48253 | 36410 |
| Unclassified fraction | 32% | 25% | 20% | 15% | 11% | 8% | 3% | -4% | -16% | 12% |
| Relative classification | under | under | under | under | under | under | under | over | over | under |

*3.4.2 Stochastic 3D Hālawa-Moanalua aquifer features*

In this section, selected aquifer features are evaluated at specified percentile levels and compared with published geologic maps and hydrogeologic observations. Both categorical and continuous variables are examined using values extracted at representative percentiles to assess agreement between sampled observations and model predictions. As an illustrative example, sampled surface geology is compared with geologic units estimated from the conceptual groundwater model (Fig. 10), which contrasts the field-sampled surface geology (Fig. 10a) with the estimated conceptual geologic model at the 70th percentile (Fig. 10b). The field-sampled geologic maps, compiled from Hunt (1996) and Sherrod et al. (2021), are published as aggregated units: the upper section includes pāhoehoe, ʻaʻā, and alluvium; the middle section includes tuff, caprock, and alluvium; and the lower section includes caprock and fill. Overall, the field-sampled surface geology is broadly consistent with the CopulaGAN-estimated surface geologic units

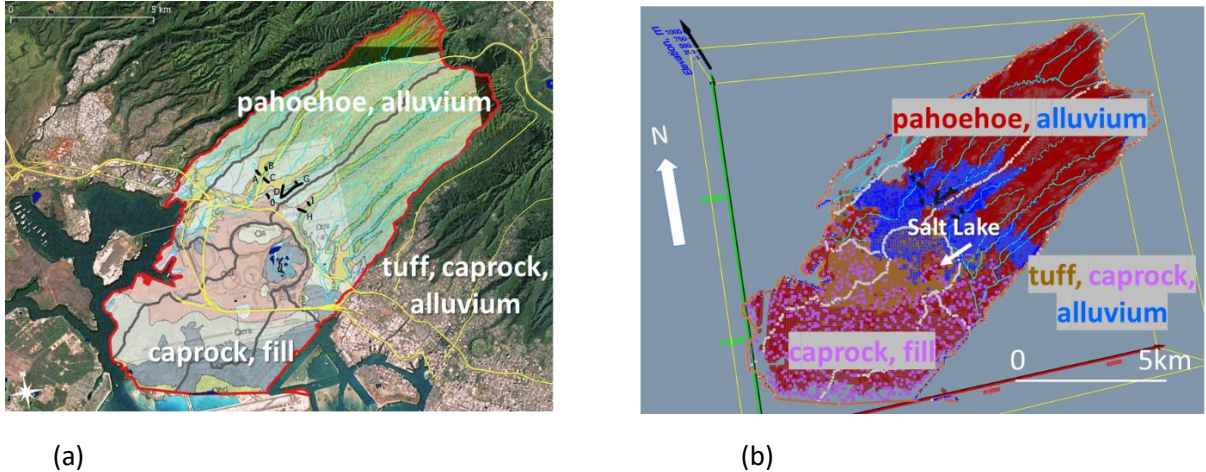

(a)  (b)

**Fig. 10.** Comparison of field sampled and estimated surface geologic units at the Hālawa-Moanalua aquifer. (a) Published field sampled surface geologic maps (Hunt, 1996; Sherrod et al., 2021). (b) CopulaGAN estimated surface geologic units (60th percentile). Caprock (magenta) reflects 1% of the available to visualize alluvium (blue), tuff (brown) and pāhoehoe (red).

In a related analysis, CopulaGAN-derived estimates of geologic units across the Hālawa–Moanalua aquifer are shown at the 50th, 60th, 70th, and 80th percentiles (Fig. 11). For clarity, alluvium (royal blue) is excluded, and caprock (magenta) is displayed at a reduced density corresponding to a random 1% subset of the total estimated locations. Visual inspection of these plots suggests a systematic increase in the spatial extent of several geologic units with increasing percentile, most notably saprolite (lime green) and tuff (brown). The expanded spatial extent of tuff at higher percentiles corresponds to delineation of the Salt Lake feature in the eastern portion of the tuff cone. Although the overall locations of the geologic unit features generally appear consistent with existing geologic interpretations, the occurrence of saprolite within the Salt Lake cone deviates from current understanding. This feature becomes more pronounced at the 80th percentile and is therefore interpreted as a modeling artefact, potentially related to limitations in the one-hot encoding assignment and/or the SOM mapping parameters.

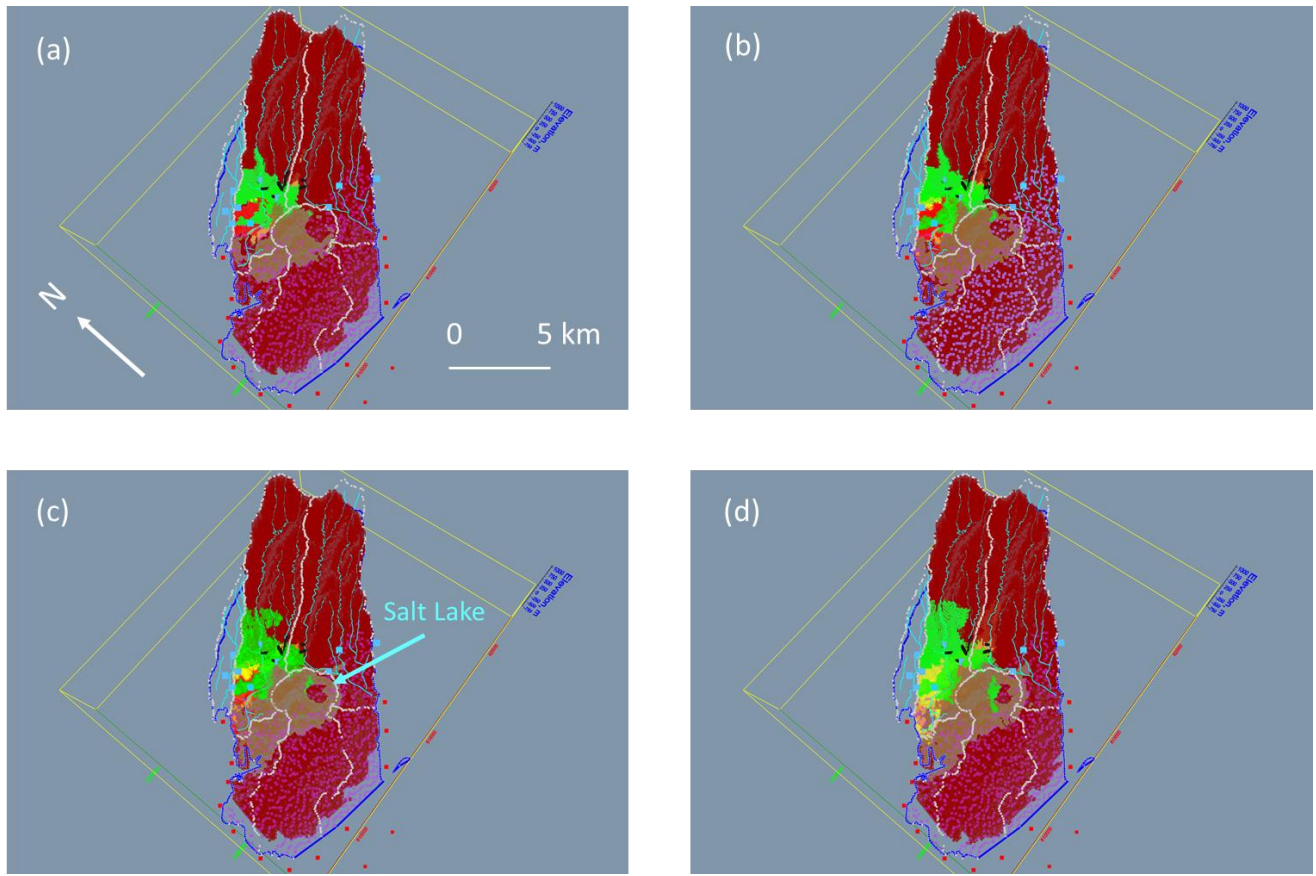

**Fig 11.** Top view of the CopulaGAN 3D geologic unit estimates (fake) across the Hālawa-Moanalua aquifer. Units: ʻaʻā (orange), alluvium (royal blue), caprock - 1% of total (magenta), saprolite (lime green), clinker-loose (yellow), clinker-welded (red), pāhoehoe (brick red), tuff (brown), (a) 50$^{th}$ percentile, (b) 60$^{th}$ percentile, (c) 70$^{th}$ percentile, (d) 80$^{th}$ percentile. Notice that the area of geologic units enlarges with increasing percentile. The alluvium is removed to highlight spatial changes in the geologic unit predictions. The black lines appearing in the saprolite (green) indicate the locations of seismic refraction and reflection profiles, and blue squares are supply wells. The location of Salt Lake appears at the center of the tuff cone directly overlying the undifferentiated pāhoehoe lava.

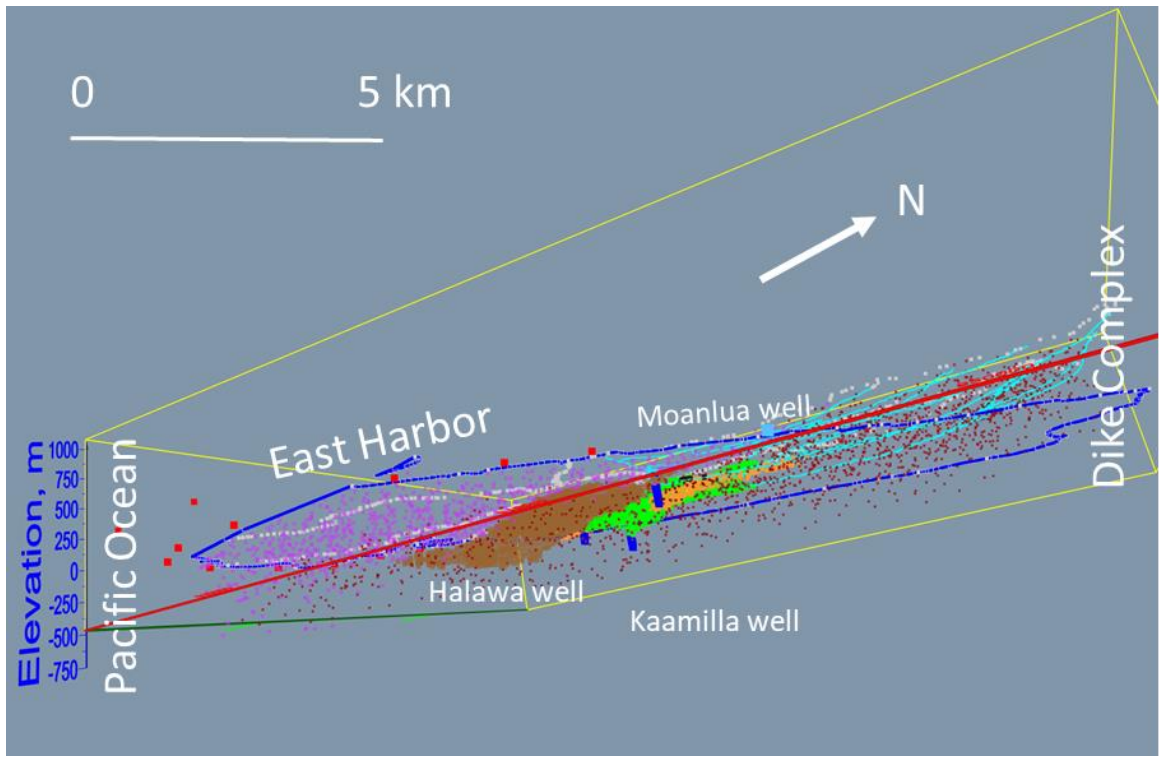
(a)

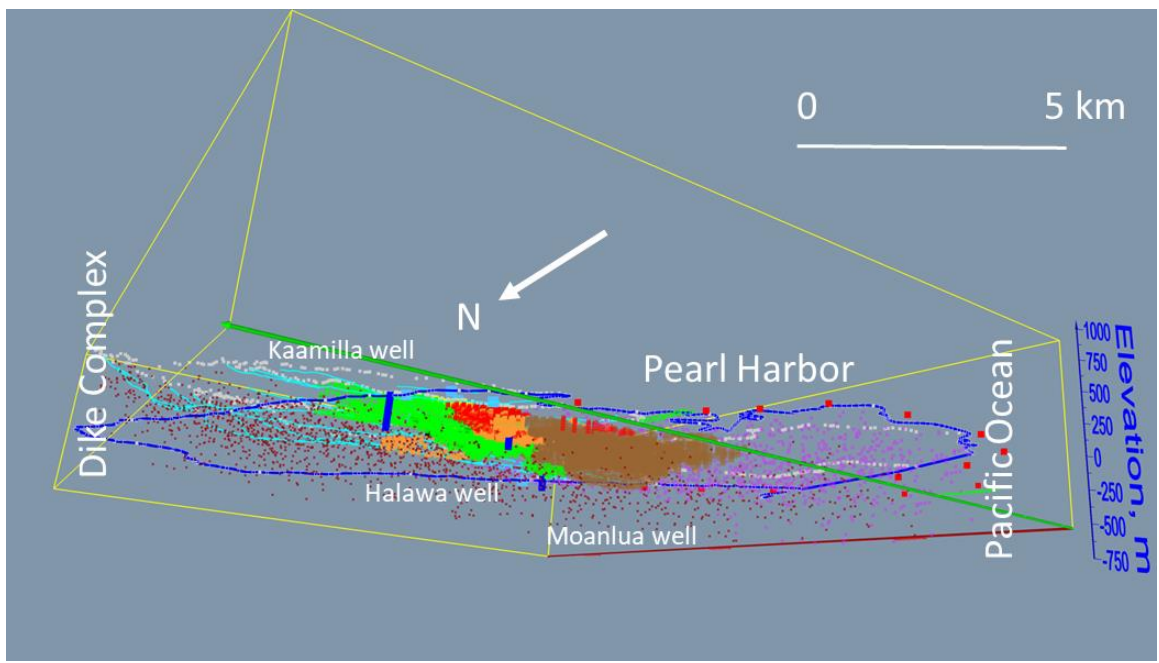
(b)

**Fig. 12.** Side-view perspective looking upward toward the estimated (synthetic) geologic units across the Hālawa–Moanalua aquifer. Geologic units include ʻaʻā (orange), caprock (magenta; ~1% of total), saprolite (lime green), clinker–loose (yellow), clinker–welded (red), pāhoehoe (brick red, ~1% of total),

and tuff (brown). (a) East–west view. (b) West–east view. Vertical blue lines are deep monitoring wells. Red squares are harbor nodes overlying seawater.

Field-sampled aquifer conditions are compared with TVAE-estimated conditions at the 70th percentile in Fig. 13. This comparison provides an independent assessment of the CGM's ability to reproduce the spatial distribution of the unconfined/unsaturated zone (white) and the confined/saturated zone (dark blue). Reference data include depth-to-water measurements (light blue squares) collected at monitoring well locations (red squares), surface projections of seismic profiles (black lines), and area monitoring wells (red squares). The figure panels include: (a) a plan view showing water-related features (Fig. 13a), (b) a cross-sectional view illustrating water-related features (Fig. 13b), and (c) a plan view combining water and geologic unit features (Fig. 13c), including clinker welded (dark orange), clinker loose (yellow), ʻaʻā lava (orange), saprolite (bright green), and alluvium overlying the seismic profiles (royal blue).

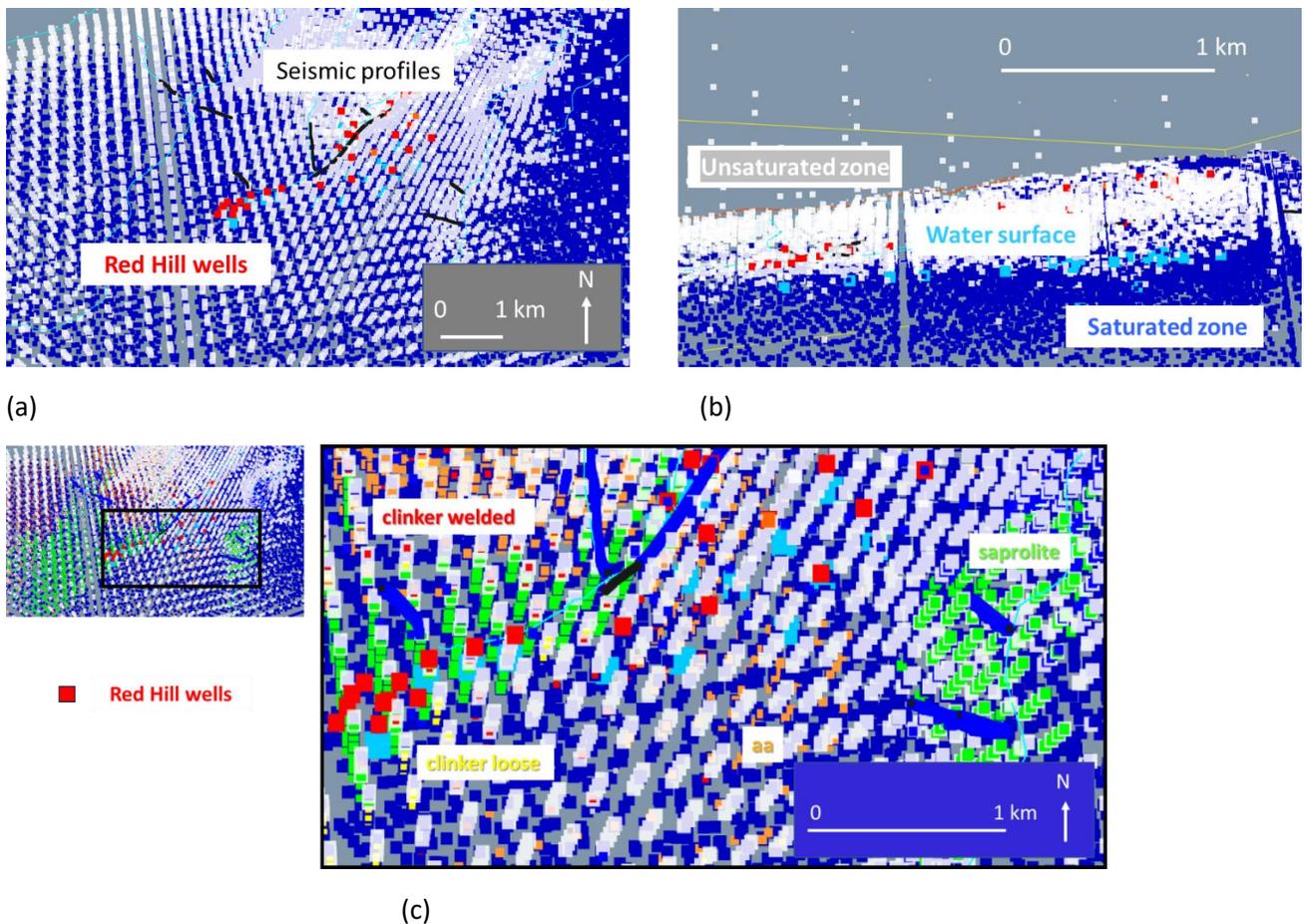

(a) (b)

(c)

**Fig. 13**. Comparison of field-sampled aquifer conditions with TVAE-estimated conditions at the 70th percentile. Panels show (a) plan view of unconfined/unsaturated (white) and confined/saturated (dark blue) zones with reference water data, (b) cross-sectional view of water features, and (c) plan view combining water and geologic unit features. Reference data include depth-to-water measurements at monitoring wells (light blue squares), monitoring well locations (red squares), and surface projections of seismic profiles (black lines). Geologic units include clinker welded (dark orange), clinker loose (yellow), ʻaʻā lava (orange), saprolite (bright green), and alluvium overlying seismic profiles (royal blue).

Next, aquifer temperature and conductance were evaluated at the 50th and 75th percentiles to examine spatial variability and overall model behavior. Overall, the estimated temperature and conductance values show good agreement with observations from the deep Hālawa and Moanalua monitoring wells (Board of Water Supply, 2023). This agreement suggests that the modeled distributions provide a reasonable representation of subsurface conditions across the study area. Differences between percentiles are generally small and localized, indicating limited sensitivity of the model to the chosen percentile.

Minor temperature differences between the 50th and 75th percentiles are observed in the lower half of the Hālawa well and extend laterally toward the Moanalua well. These differences correspond to temperature increases of approximately 0.5 °C, as indicated by a color shift from royal blue to green in the percentile comparison plots. Despite these localized variations, the overall temperature structure remains consistent between percentiles, demonstrating stability in the modeled thermal profile across the aquifer.

Similarly, minor changes in conductance are evident between the gridded percentile estimates. Slight increases occur beneath the Hālawa and Moanalua wells and east of the Moanalua well when moving from the 50th to the 75th percentile. Conductance values rise from approximately 5,000 µS/cm, indicative of freshwater, to about 12,500 µS/cm, representative of brackish conditions, with the changes reflected by a color shift from royal blue to green in the plots. These variations are localized but consistent with the overall spatial trends observed in the aquifer.

Farther east in the model domain, conductance also increases when comparing the 50th and 75th percentile plots. Overall, the spatial patterns of temperature and conductance remain very similar between percentiles and closely match the estimated conditions surrounding the Hālawa and Moanalua wells. This consistency suggests that the choice of percentile does not substantially alter the interpretation of aquifer conditions. However, because some grid nodes remain unclassified at the 50th percentile while all nodes are classified at the 75th percentile (Table 11), the following section focuses on interpreting three-dimensional conductance plots in the context of freshwater, seawater, and their exchange.

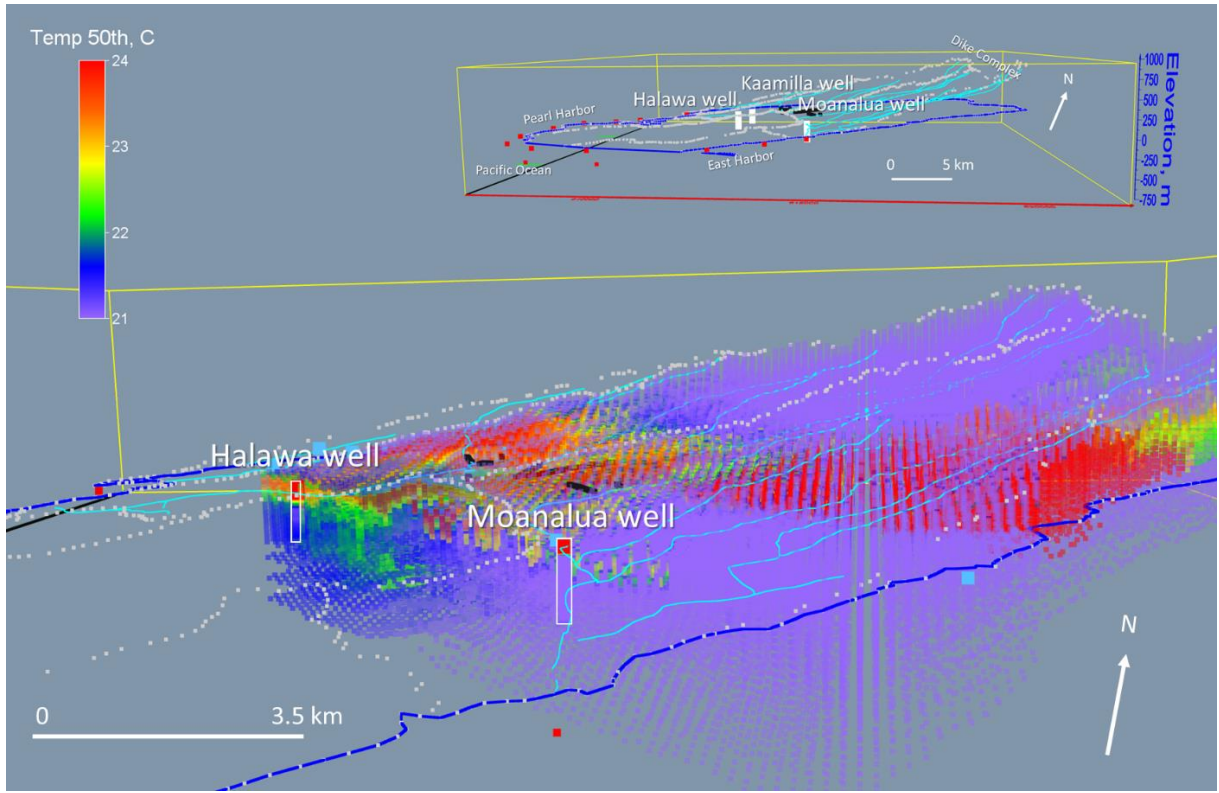

(a)

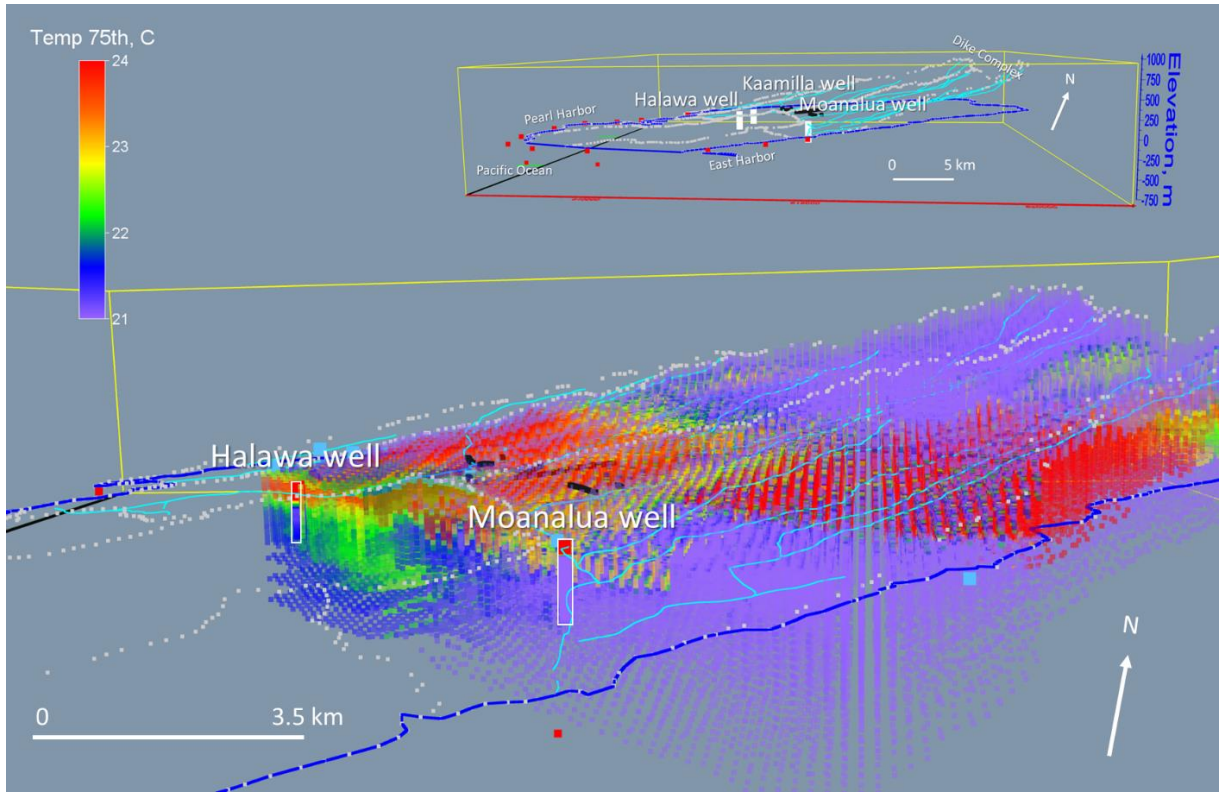

(b)

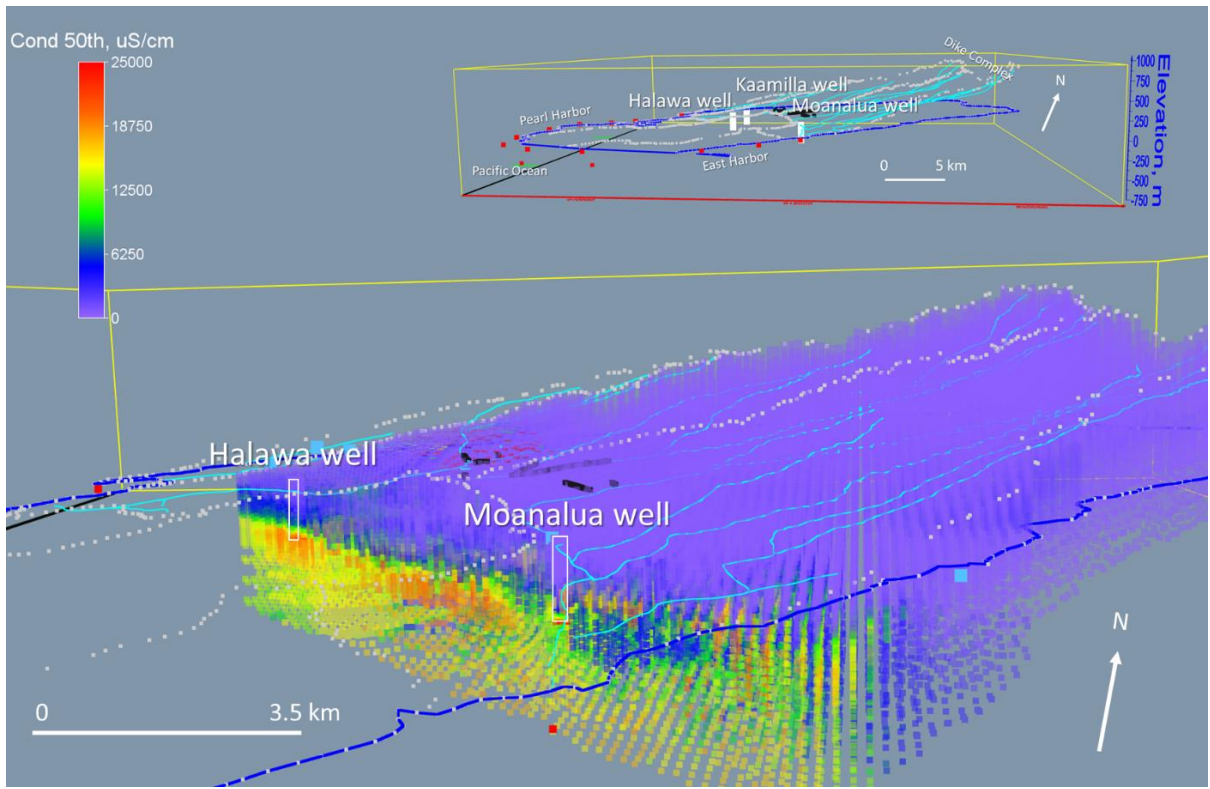

(c)

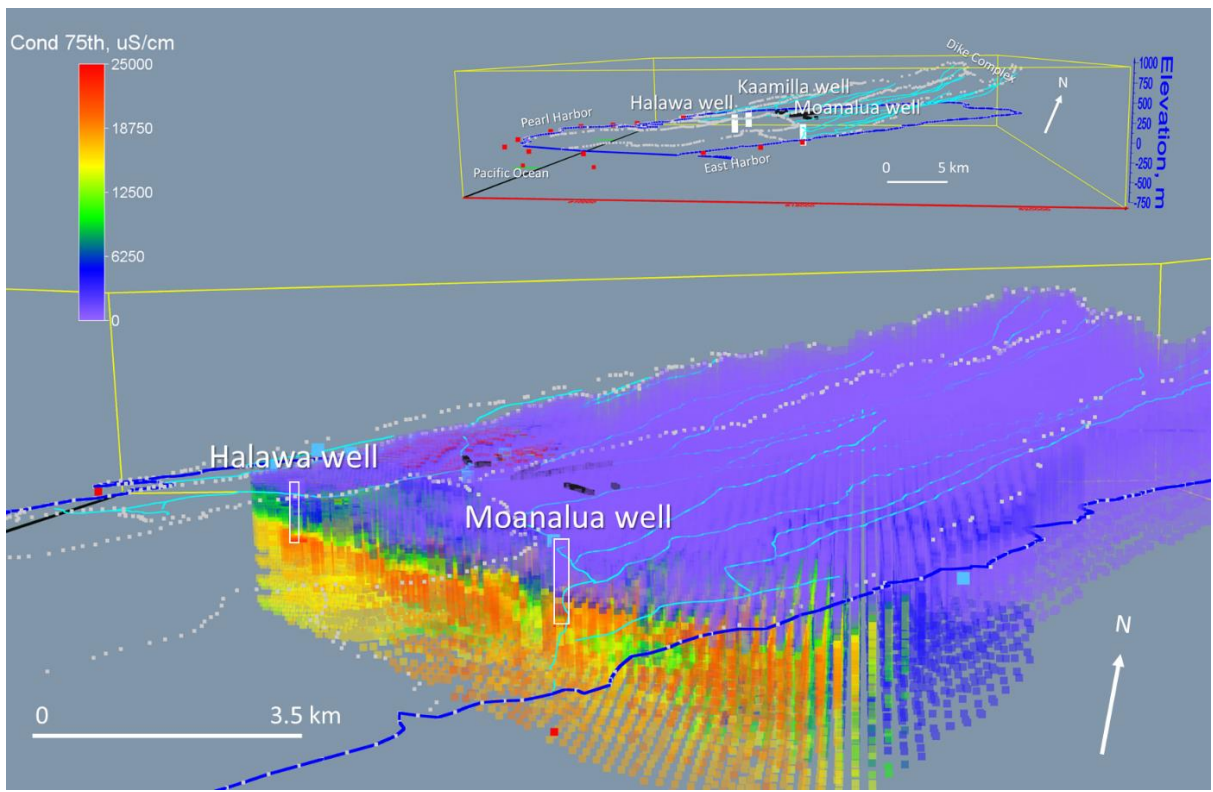

(d)

**Fig 14.** CopulaGAN-derived estimates of aquifer temperature and conductance compared with measured profiles from the Hālawa and Moanalua deep monitoring wells (outlined in white). Panels show (a) median (50th percentile) temperature, (b) 75th percentile temperature, (c) median (50th percentile) conductance, and (d) 75th percentile conductance. Model grid surface elevations and depths at the Hālawa and Moanalua wells are approximately 18.1 m / 650 m and 10.8 m / 650 m, respectively. The inset shows the locations and depths of the monitoring wells relative to the Hālawa and Moanalua aquifer. Streams are indicated by light blue lines, and ocean nodes by red squares.

The three-dimensional groundwater conductance model is evaluated using the 75th percentile of CopulaGAN continuous-feature estimates. On-land conductance values across the Hālawa–Moanalua aquifer span nearly four orders of magnitude, from ~100 µS/cm (low-conductance freshwater) to ~25,000 µS/cm (high-conductance saline water, ~50% of seawater conductivity). This range provides a quantitative basis for delineating freshwater, brackish, and saline zones, as well as areas of freshwater–seawater exchange, reflecting lithologic heterogeneity, structural controls, and hydraulic gradients.

Low-conductance groundwater associated with the northern dike complex flows downgradient toward the major supply wells. Freshwater exhibits density-dependent behavior near the wells, particularly along upslope aquifer margins and north of both harbors (Figs. 15a, c). Pumping appears to capture both dike-derived freshwater and brackish water from adjacent mixing zones, consistent with lateral inflow and vertical density stratification. This suggests that well operations influence local salinity distributions and enhance freshwater–brackish water interactions.

Freshwater discharge is inferred along the western aquifer margin, where preferential pathways through undifferentiated pāhoehoe units appear to convey groundwater directly to the Pacific Ocean below Pearl Harbor (Fig. 15b). These pathways, oriented roughly perpendicular to the coastline, bypass the supply wells and likely represent a previously underrecognized component of the regional water budget. East of this discharge zone, seawater intrusion appears to occur along parallel, depth-dependent pathways, indicating structural or lithologic control on flow.

Conductance at the base of the aquifer supports freshwater north of the wells, with values increasing downgradient toward the ocean. The vertical transition from freshwater to brackish and saline water is consistent with coastal mixing processes. Highest conductance occurs beneath the Salt Lake Tuff cone (Fig. 11c), where pāhoehoe replaces the tuff cone center, providing a preferential pathway for seawater intrusion. The spatial coincidence of elevated conductance and lithologic change supports focused seawater transport and discharge.

Across uplands and ridges, conductance varies, but it remains relatively uniform beneath river channels (Fig. 15d), suggesting limited hydraulic connection between rivers and the basal aquifer. Potentiometric surfaces (Fig. 12) further support this interpretation. Conductance values consistent with seawater (~44,000 µS/cm) are observed only at harbor nodes, indicating localized intrusion.

Beyond conductance, the CGM framework can represent additional continuous features, including geophysical (velocity, density), engineering (fracture rate, rock quality), and water-quality parameters (dissolved oxygen, redox potential, pH). Hydraulic properties governing flow (hydraulic conductivity, specific storage, specific yield) and transport (bulk density, dispersivity, porosity) can also be mapped to CGM grid nodes via the SOM. This approach will enable application of boundary conditions and calibration of numerical models for simulating freshwater–seawater exchange within the Hālawa–Moanalua aquifer.

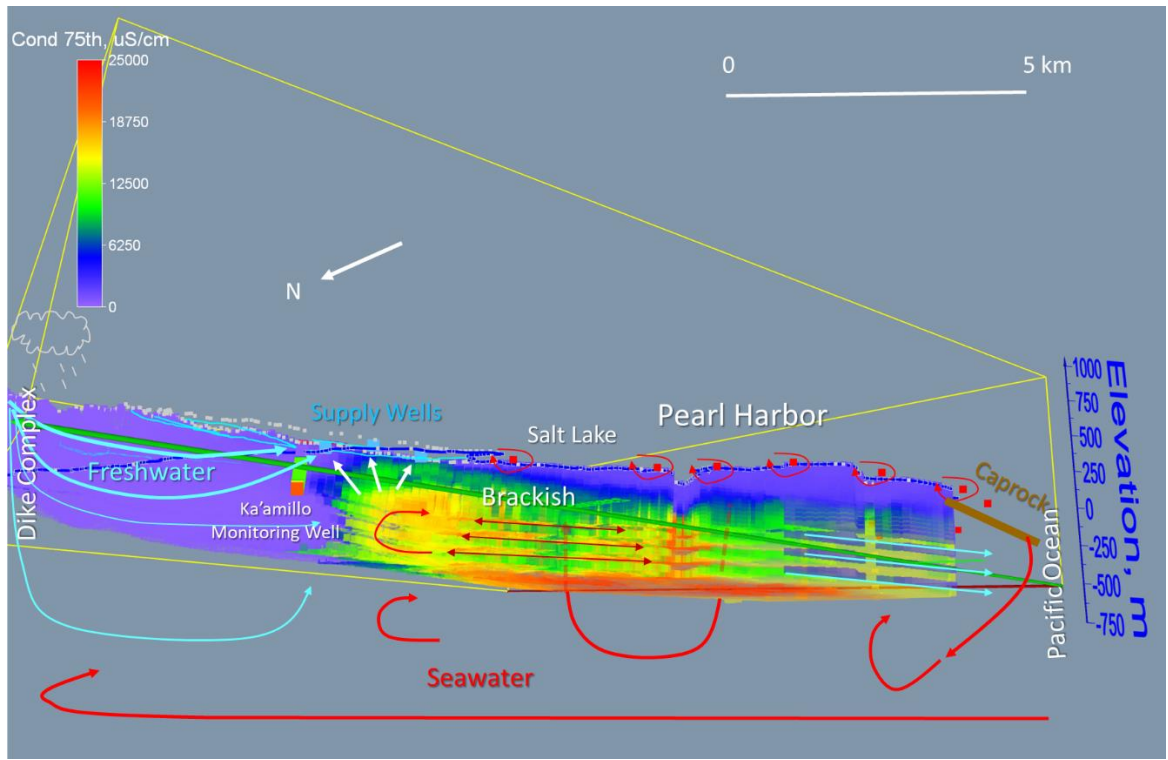

(a)

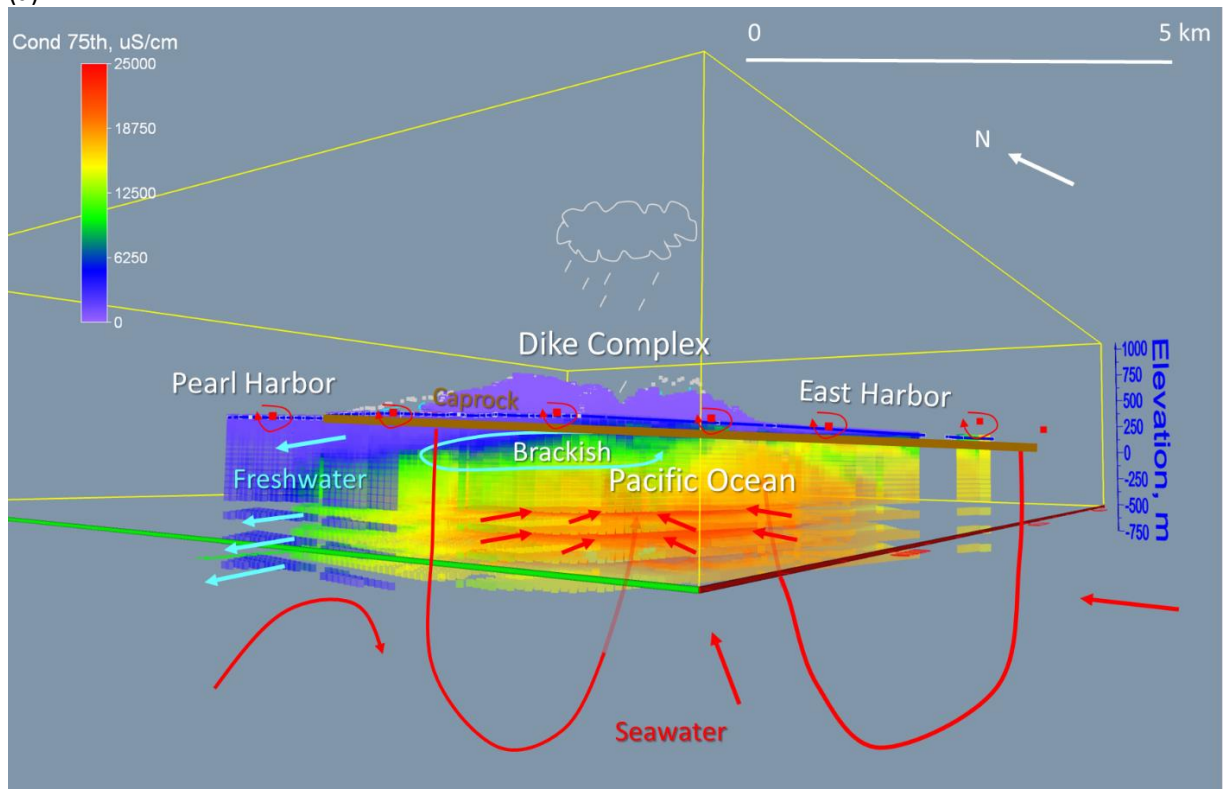

(b)

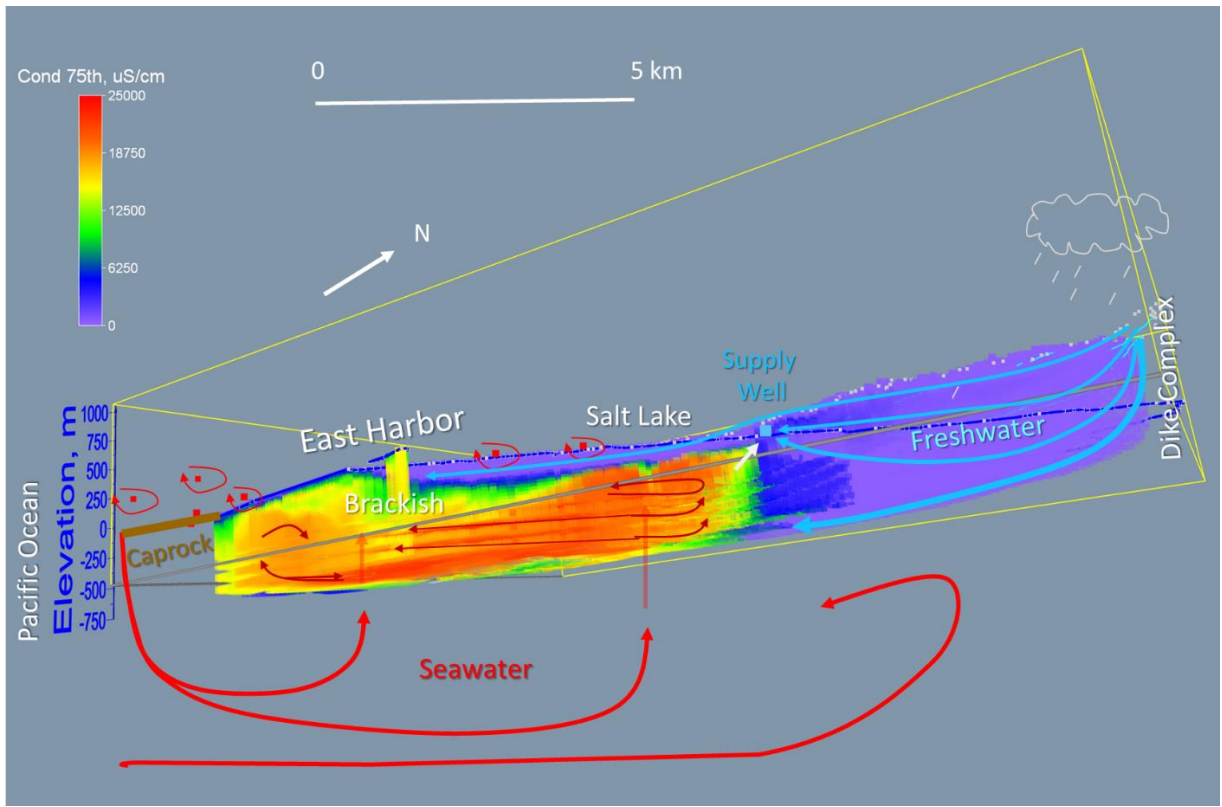

(c)

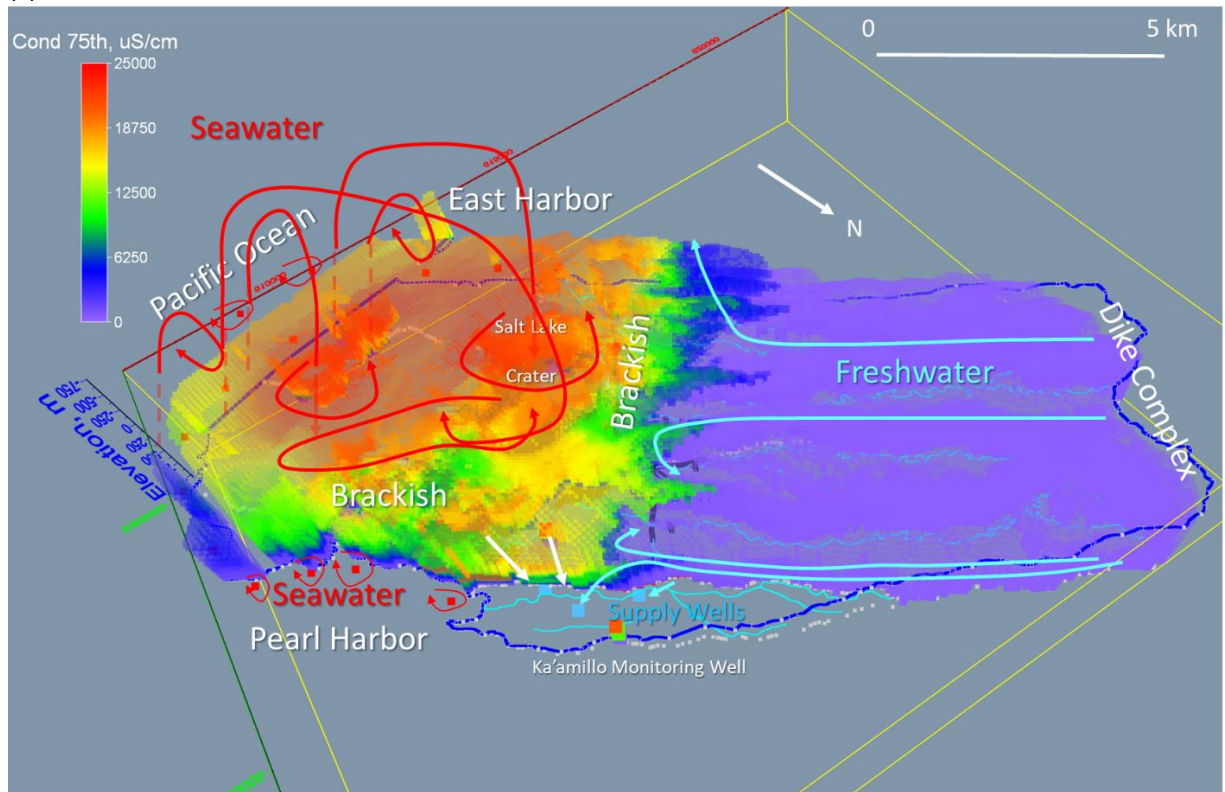

(d)

**Fig 15.** CopulaGAN estimates of conductance (Cond) revealing freshwater-seawater exchange across the Hālawa-Moanalua aquifer. (a) west to east view, (b) north to south view, (c), east to west view, and (d) bottom to top view (dike complex is along the northern boundary and Pacific Ocean along the southern boundary). The red squares represent nodes located in both harbors and the Pacific Ocean.

## 4. Conclusions

This study demonstrates that integrating unsupervised machine learning with generative artificial intelligence provides a robust framework for characterizing Pacific volcanic groundwater systems using sparse and highly imbalanced data. Among the methods evaluated, the Copula Generative Adversarial Network (CopulaGAN) was statistically preferred for developing the conceptual groundwater model (CGM) of the Hālawa–Moanalua aquifer, Oʻahu, Hawaiʻi. For groundwater-level estimation, the Tabular Variational Autoencoder (TVAE) performed best, producing geologic unit representations comparable to those generated by CopulaGAN and the Tabular Gaussian Copula. These results highlight the value of applying multiple generative approaches to improve CGM robustness for complex coastal volcanic aquifers.

CopulaGAN-generated three-dimensional aquifer features reproduce published surface geologic maps and show strong agreement with observed conductance, temperature, and barometric pressure profiles. Conductance models indicate groundwater flow and discharge are influenced by regional hydraulic gradients, pumping, and seawater intrusion, as well as preferential pathways for freshwater–seawater exchange. Key hydrogeologic processes captured by the models include freshwater transport from dike complexes through pāhoehoe lava to slope transitions, submarine discharge along linear pathways to the Pacific Ocean, and landward seawater intrusion along southern and deep flow paths toward the Salt Lake crater.

Beyond conductance, the CGM framework can incorporate additional geophysical, engineering, and water-quality parameters, as well as hydraulic and transport properties, enabling future mapping, calibration, and simulation of aquifer dynamics. Overall, the CopulaGAN approach demonstrates strong potential for integrating sparse data, resolving aquifer heterogeneity, and supporting quantitative assessments of freshwater–seawater exchange. These findings provide critical insight for water-resource management, highlighting zones of active mixing, preferential flow, and localized seawater intrusion that may influence both supply-well performance and regional groundwater sustainability. The workflow is transferable to other Pacific-island aquifers and broader subsurface applications in the energy, mineral, and water-resource sectors.

**CRediT authorship contribution statement**

**Michael J. Friedel**: Writing – Original draft, Review & editing, Software, Methodology, Conceptualization, and Funding acquisition.

**Declaration of competing interest**

The authors declare that they have no known competing financial interests or personal relationships that could have appeared to influence the work reported in this paper.

**Acknowledgments**

This research was supported by a U.S. Navy grant (contract # 6112062-02) during the authors' appointment as Associate Researcher at the Research Corporation of the University of Hawaiʻi. The author gratefully acknowledges Drs. Donald Thomas, Erin Wallin, Peter Kannberg, and Toomas Parratt for their valuable support and contributions.

**Data availability**

Data will be made available on request.